\documentclass{aa}  
\usepackage{graphicx}
\usepackage{txfonts}
\usepackage{caption}
\usepackage{hyperref}
\usepackage{natbib}
\bibpunct{(}{)}{;}{a}{}{,} 

\begin{document} 
\title{The origin of dust polarization in molecular outflows}
\author{S. Reissl\inst{\ref{inst1},\ref{inst2}} \and D. 
Seifried\inst{\ref{inst3}} \and S. Wolf \inst{\ref{inst2}} \and R. Banerjee 
\inst{\ref{inst4}} \and R. S. Klessen \inst{\ref{inst1}}}

\institute{
\centering University of Heidelberg, Institute of Theoretical Astrophysics, 
Albert-Ueberle-Str. 2 U04, 69120 Heidelberg, Germany \\				
\href{mailto:reissl@uni-heidelberg.de}{reissl@uni-heidelberg.de},
\href{mailto:klessenl@uni-heidelberg.de}{klessen@uni-heidelberg.de}\label{inst1}
\and
\centering Institut für Theoretische Physik und Astrophysik, Christian-Albrechts-Universit\"{a}t zu Kiel, Leibnizstraße 15, 24098 Kiel, Germany \\
								\href{mailto:sreissl@astrophysik.uni-kiel.de}{sreissl@astrophysik.uni-kiel.de},  
								\href{mailto:wolf@astrophysik.uni-kiel.de}{wolf@astrophysik.uni-kiel.de}\label{inst2}
\and
					 \centering I. Physikalisches Institut, Universität zu Köln, Zülpicher Straße 77, 50937 Köln, Germany \\
								\href{mailto:seifried@ph1.uni-koeln.de}{seifried@ph1.uni-koeln.de}\label{inst3}
\and
					 \centering Hamburger Sternwarte, Universität Hamburg, Gojenbergsweg 112, 21029 Hamburg, Germany \\
								\href{mailto:banerjee@hs.uni-hamburg.de}{banerjee@hs.uni-hamburg.de}\label{inst4}
}
						
\abstract
   {}
{Polarization measurements of dust grains aligned with the magnetic field 
direction are a established technique to trace large-scale field structures. In 
this paper we present a case study to investigate conditions necessary to detect 
a characteristic magnetic field substructure embedded in such a large-scale 
field. A helical magnetic field with a surrounding hourglass shaped field is 
expected from theoretical predictions and self-consistent magnetohydrodynamical (MHD)  simulations to be 
present in the specific case of protostellar outflows. Hence, such a outflow 
environment is the perfect environment for our study.
}
{We present synthetic polarisation maps in the infrared and millimeter regime of simulations of protostellar outflows performed with the newly developed radiative transfer and polarisation code POLARIS. The code, as the first, includes a self-consistent description of various alignement mechanism like the imperfect Davis-Greenstein (IDG) and the radiative torque (RAT) alignment.
We investigate for which effects the grain size distribution, inclination, and applied alignement mechanism have.
}
   {We find that the IDG mechanism cannot produce any measurable polarization degree ($\geq 1\ \%$) whereas RAT alignment produced polarization degrees of a few $1\ \%$.
Furthermore, we developed a method to identify the origin of the polarization. We show that the helical magnetic field in the outflow can only be observed close to the outflow axis and at its tip, whereas in the surrounding regions the hourglass field in the foreground dominates the polarization. Furthermore, the polarization degree in the outflow lobe is lower than in the surroundings in agreement with observations.
We also find that the orientation of the polarization vector flips around a few $100\ \rm{\mu m}$ due to the transition from dichroic extinction to thermal re-emission. Hence, in order to avoid ambiguities when interpreting polarization data, we suggest to observed in the far-infrared and mm regime. The actual grain size distribution has only little effect on the emerging polarization maps.
Finally, we show that with ALMA it is possible to observe the polarized radiation emerging from protostellar outflows. 
}
 {}
  \maketitle
%

\section{Introduction}
Magnetic fields are responsible for protosteller outflows one of the most 
prominent signs of star-formation \citep[][]{Pudritz1983}.
Such outflows are more easily to detect than the 
embedded protostellar disks, from where they are expected to be launched. 
Hence, the observation of an outflow is a strong indicator for the presence of 
a 
rotating disk. Moreover, the launching 
mechanism of the outflow is closely linked to the disk rotation since the 
magnetic field lines are frozen at the disk scale and dragged along the 
direction of  rotation. The magnetic field morphology in this scenario is 
expected to have a strong toroidal component 
\citep[e.g.][]{Blandford1982,Pudritz1983,Shibata1985,Tomisaka1998,
Banerjee2006} forming a helical field 
morphology together 
with the large-scale poloidal field of the surrounding medium. 
Hence, a detection of a helical 
field component in an outflow would be an additional indicator for the 
presence of a rotating disk embedded in the center.\\
From an observational point of view it needs to be verified by proper modeling 
if the helical field morphological would actually be detectable in the 
interior of outflow lobes or whether the surrounding hourglass-field dominates 
light polarization. While recent observations of CO polarization 
measurements seem to confirm the hypothesis of a helical field 
\citep[][]{2016ApJ...819..159C}, the interpretation of such data 
remains still unclear since the magnetic field morphology in the ambient 
environment represents a possible source of ambiguity. The large-scale  
magnetic field is expected to be hourglass shaped and is therefore orientated
perpendicular to the toroidal component along the 
line-of-sight (LOS) hiding the embedded helical field morphology 
\citep[e.g.][]{Girart2006,Girart2009}. \\
Focusing on the aspect of the magnetic morphology this can be 
probed by dust polarization measurements. Historically, the alignment of 
non-spherical dust grains has long been suggested as a explanation 
for the observed polarization of stellar radiation 
\citep[][]{1949Natur.163..283H,1949Sci...109..166H,1971MNRAS.153..279M}. \\
Different theories of grain alignment agree that rotating non-spherical dust 
grains tend to align with the magnetic field direction 
\citep[see][]{2007JQSRT.106..225L}. Unpolarized light will gain polarization by 
dichroic extinction in the mid-infrared (mid-IR) and thermal dust 
re-emission from far-infrared (far-IR) to sub-millimeter (sub-mm) and 
millimeter 
(mm) wavelength \citep[e.g.][]{2011AA...535A..44F} making it possible to infer 
the projected magnetic field morphology. Hence, polarization measurements 
provide a promising tool to determine the morphology and 
subsequently to investigate of the role of  magnetic fields in the 
evolution 
\citep[][]{2000PASP..112.1215H,2004mim..proc..123C,2012ApJ...751L..20G} of 
star-forming systems. The difficulty is that the reliability of polarization 
measurements and their interpretation depends on a wide range of physical 
parameters that are still discussed \citep[see][for 
review]{Andersson2015}.\\ 
With the dedicated instruments such as the Atacama 
Large Millimeter Array  
\cite[ALMA,][]{2004AdSpR..34..555B}, and the high-resolution Airborne Wideband 
Camera-plus instrument of the airborne Stratospheric Observatory For Infrared 
Astronomy \cite[HAWC+/SOFIA,][]{Dowell2013}, the measurement of dust 
polarization of cloud cores and disks potentially becomes feasible. Hence, 
questions about the potential of multi-wavelength polarization measurements to 
identify specific components of larger structures with complex magnetic field 
morphology.\\
Indeed, there are already a number of observations measuring the magnetic field 
of molecular outflows and the magnetic field in the center of molecular cloud 
cores. The results, however, seem to contradict each other. Whereas 
\citet{Davidson11} and \citet{Chapman13} report magnetic fields which are 
preferentially aligned with outflows, \citet{Hull13,Hull14} find magnetic 
fields to be strongly misaligned with respect to the outflow axis. Hence, 
simulations of synthetic observations are essential to asses to what accuracy 
the structure of magnetic fields in star forming cores can be inferred from 
actual observations. Even for a well-defined field structure as present in 
\cite{Reissl2014} the resulting polarization pattern is extremely 
complex and thus not easy to interpret. This holds even more in case that 
turbulent motions are involved. Hence, we argue that possible conclusions drawn 
from such observations have to be considered with great care.\\
The problems tied to the correct interpretation of dust 
polarization observations that need to be addressed are:
\begin{itemize}
\item The measured polarization will naturally only be a projection of the 
underlying magnetic 
field morphology since it is averaged along a particular LOS. The question 
remains to what extend - if at all - the 3D structure of the underlying 
magnetic field can be deduced by dust polarization measurements.
  \item It is not clear a priori whether the polarization is observed in 
dichroic extinction or re-emission since both competing mechanisms are  at work 
simultaneously. This ambiguity whether the projected magnetic field is 
perpendicular or parallel to the measured polarization vector is often neglected 
in the literature.
  \item The observed degree of polarization strongly depends on the composition 
and size of the dust grains. Hence, uncertainties emerge from unconstrained 
dust properties, in particular in dense star forming regions.
  \item  There are 
different theories which can account for the alignment of dust 
grains under certain conditions. Each dust alignment theory comes with its 
characteristic polarization pattern making the analysis of polarization 
measurements highly dependent on the choice of the considered theory.
\end{itemize}
In order to claim a correlation between the observed orientation of linear 
polarization and a particular magnetic field structure, it requires careful RT 
modeling to ensure that the observed wavelengths are actually suitable to probe 
the regions of interest with aligned dust grains in the first place. However, 
for RT simulation modeling dust polarization maps often just the density 
weighted-magnetic field is added up along the LOS 
\citep[e.g.][]{2013ApJ...774..128S} or the degree of polarization is adjusted to 
match observational data \citep{2012A&A...543A..16P}. Thus, these attempts are 
of limited predictive capability since they oversimplify the complex physics of 
radiation-dust interaction and dust grain alignment physics.\\
For this reason it is required to perform fully self-consistent radiative 
transfer simulations on linear polarization of radiation incorporating dust 
grain composition and a well-motivated grain alignment efficiency. In order to 
do so, we make use of the newly developed 3D RT code POLARIS 
\citep[][]{Reissl2016}, the only code that is currently available capable of 
polarization calculations on non-spherical partially aligned dust grains. We  
study the polarization of mid-IR to mm 
radiation due to aligned dust grains. Here, we will present a first analysis of 
synthetic polarization observations 
obtained from a post-processed MHD simulation 
\citep[see][]{Seifried11,Seifried12}. The simulation models the self-consistent 
formation of a protostellar disk and its associated molecular outflow driven 
in the Class 0 stage.\\
The structure of the paper is as follows: First, we present the properties of 
the chosen MHD simulation in Sect. \ref{sect:MHD}. We then show details of the 
applied dust model in Sect. \ref{sect:DustModel}. In Sect. \ref{sect:align} the 
considered theories of dust grain alignment are introduced followed by the 
description of the RT techniques in Sect. \ref{sect:RT}. We show the 
synthetic intensity and polarization maps as a function of different RT 
simulation parameters in Sect. \ref{sec:results}. In Sect. \ref{sect:LOS} we 
present a method to determine the origin of polarization by means of tracing 
distinct rays in RT calculations. We deal with the influence of dust grain 
size on polarization in Sect. \ref{sect:GrainSize} and simulate actual 
$ALMA$ polarization observations in Sect. \ref{sect:Observation}.  Finally, we 
discuss and summarize the results in Sects. \ref{sect:disc} and 
\ref{sect:summ}, respectively.

\section{MHD - Outflow Simulation}
\label{sect:MHD}
\begin{figure*}[]
	\begin{minipage}[c]{1.0\linewidth}
			\begin{center}
			  \includegraphics[width=0.30\textwidth]{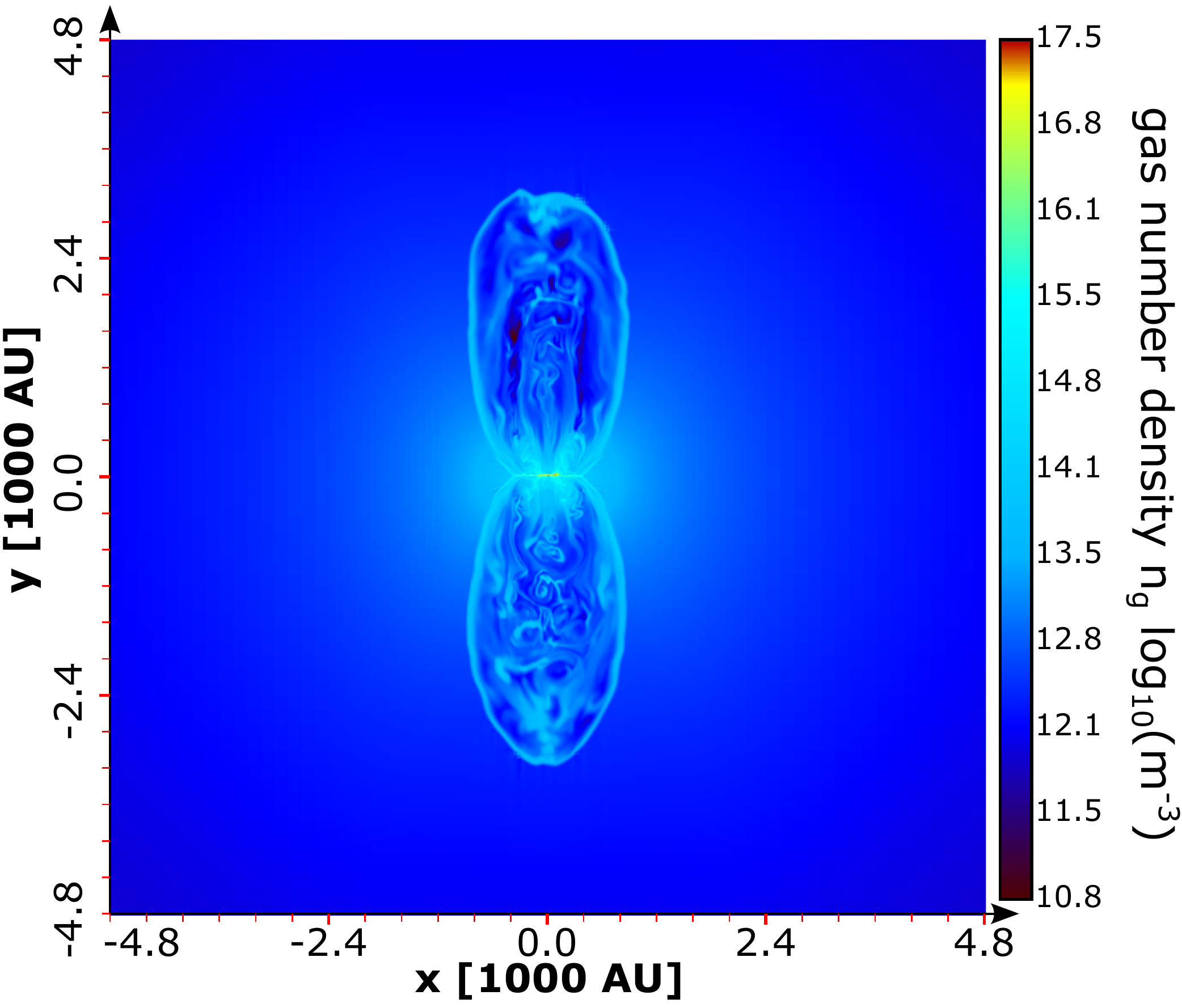}
				\hspace{5mm}
				\includegraphics[width=0.30\textwidth]{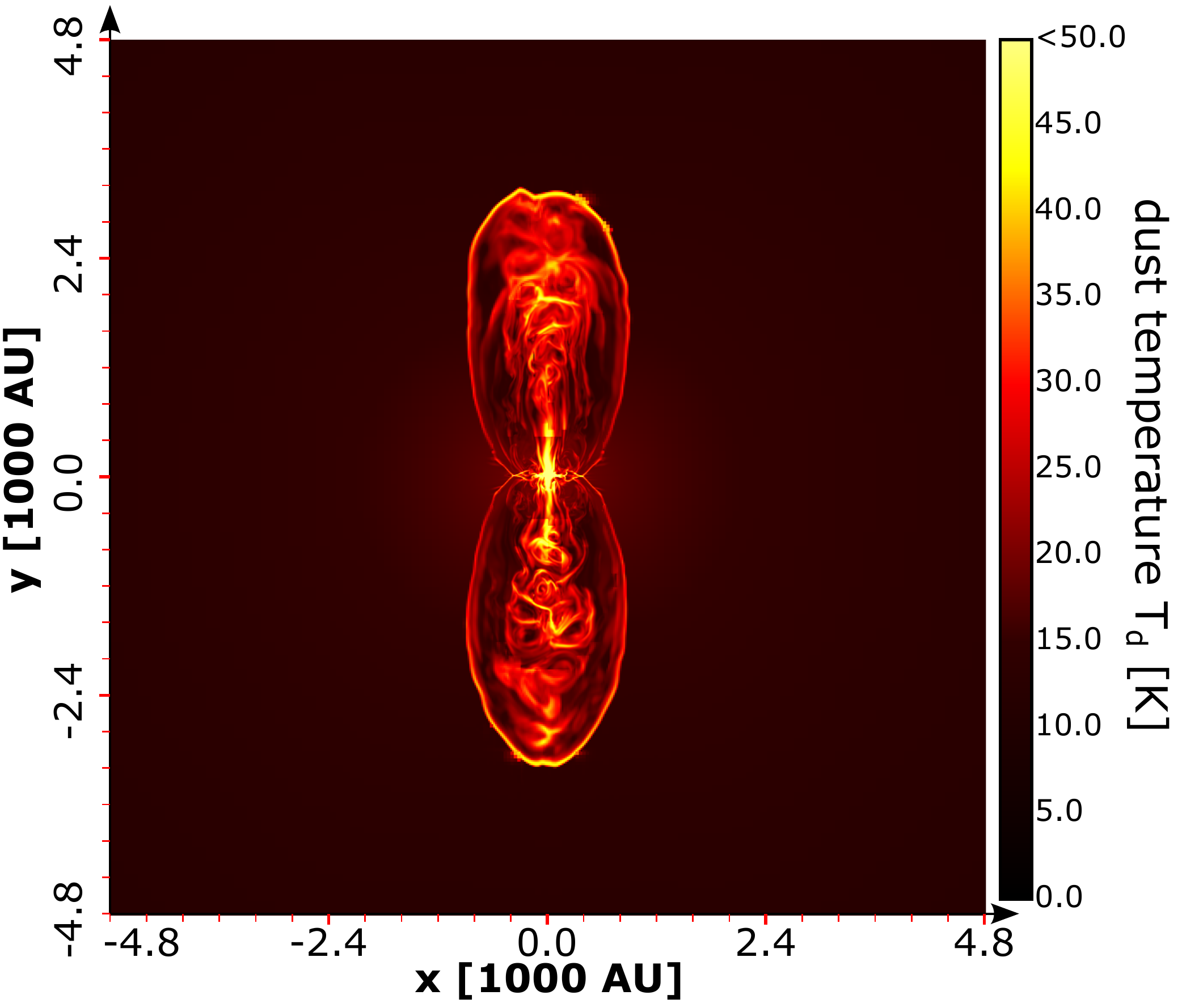}
				\hspace{5mm}
				\includegraphics[width=0.30\textwidth]{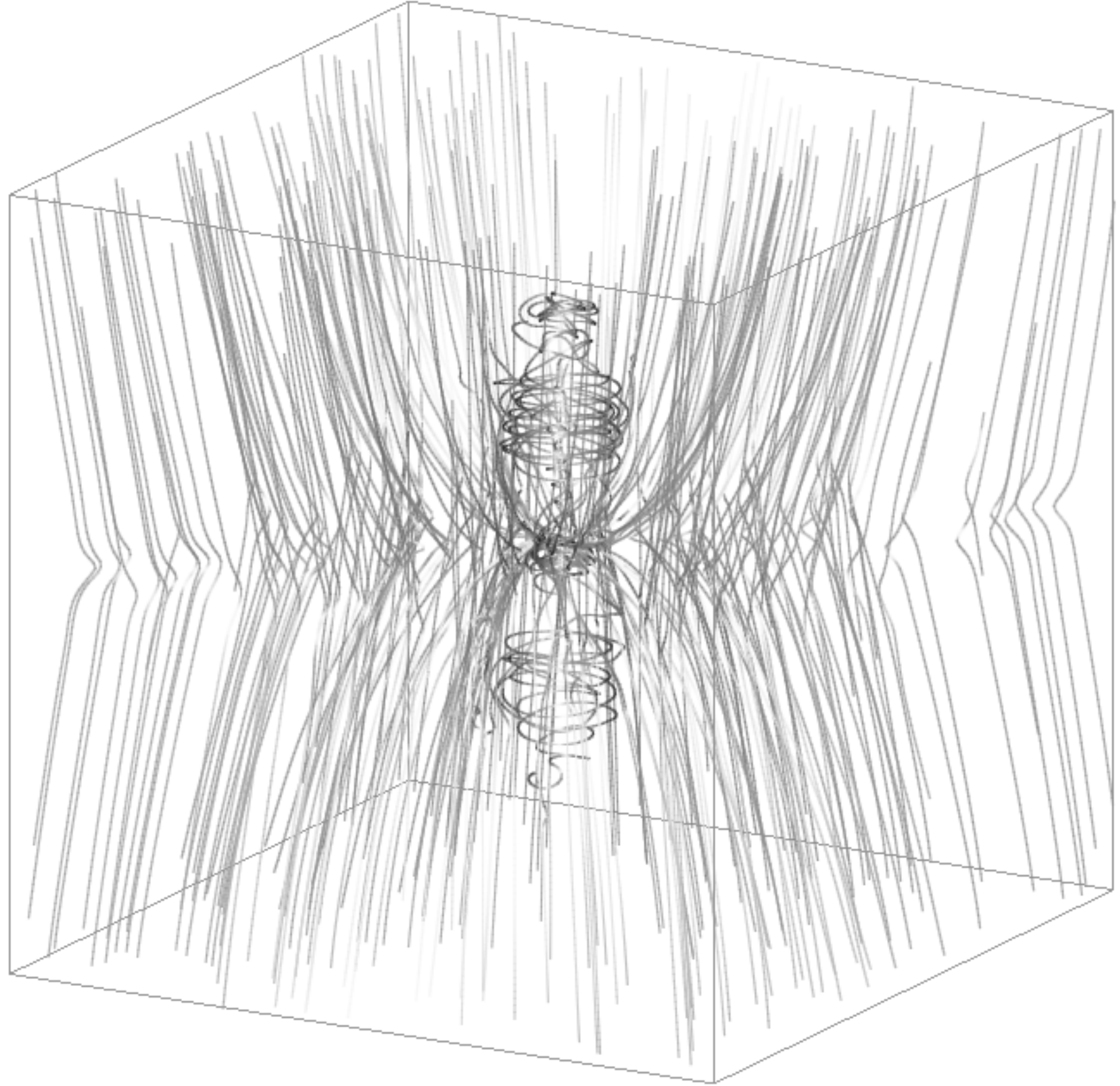}
			\end{center}
		\end{minipage}	
			
	\caption{Left panel: Gas number density $n_{\rm{g}}$ of the considered MHD  
simulation in the mid-plane parallel to the symmetry axis of the outflow lobes. Middle panel: Corresponding dust temperature $T_{\rm{g}}$. The dust temperature was post-processed to account for the stellar contribution of the protostars to dust heating (see Sect. \ref{sect:align} for details). Right panel: 3D magnetic field distribution. The characteristic helical component is in the 
interior of the outflow lobes embedded within a large scale hourglass shaped 
field. The box has an edge lengths of $9200\ \rm{AU}$.}
\label{fig:3DMag}
\end{figure*}
The MHD simulation considered in this paper results in the formation of 
protostars, their surrounding disk, and the protostaller outflow following the 
collapse of the parental protostellar core. These simulations are performed 
with the astrophysical code 
FLASH~\citep{Fryxell00} using a robust MHD solver developed by 
\citet{Bouchut07}. To follow the long-term evolution of the protostellar disk 
and its associated outflow we make use of sink particles~\citep{Federrath10}. 
For more details on the numerical methods applied we refer the reader to 
\citet{Seifried11,Seifried12}.\\
In this paper we analyze the results of a particular simulation considering 
the collapse of a 100~$M_{\sun}$ molecular cloud core without initial 
turbulence, which is $0.25\ \rm{pc}$ 
in diameter and initially rotating rigidly around the $z$-axis with a rotation 
frequency of $3.16 \cdot 10^{-13}$ s$^{-1}$. The ratio of rotational to 
gravitational 
energy is $\beta_\mathrm{rot} = 4 \cdot 10^{-2}$. The magnetic field is 
initially aligned with the rotation axis, i.e. parallel to the $z$-axis and has 
a strength chosen such that the normalized mass-to-flux ratio is 
\citep[][]{Spitzer1978}
\begin{equation}
  \mu = \left( \frac{M_\textrm{core}}{\Phi_{\textrm{core}}} 
\right)/\left(\frac{M}{\Phi}\right)_{\textrm{crit}} = 
\left(\frac{M_\textrm{core}}{\int B_z 
\mathrm{d}A}\right)/\left(\frac{0.13}{\sqrt{G}}\right) = 26 \, .
\end{equation}
This corresponds to a initial central magnetic field strength of  $132\ \rm{\mu 
G}$. Hence, the core is magnetically super-critical and the magnetic field can 
not significantly counteract the initial gravitational collapse of the core. 
Again, we refer to \citep[][ - the simulation is called 
$26-4$ within these papers]{Seifried11,Seifried12} for more details on the 
initial conditions.\\
During the collapse of the core, a rotationally supported disk builds up around 
the first protostar. The disk starts to fragment after $\sim 2600\ \rm{yr}$ 
forming a small cluster of altogether six protostars. A magneto-centrifugally 
driven 
protostellar outflow is launched after the formation of the first sink 
particle, which is well-collimated with a collimation factor of $\sim 4$ by the 
end of the simulation. The outflow continues expanding in a roughly 
self-similar fashion keeping the overall morphological properties.\\
In this work we analyze this outflow at an age of 
$5000\ \rm{yr}$. By this time the outflow lobe has 
a height of $3200\ \rm{AU}$ above and below the disk midplane and a maximum 
outflow speed of about $19\ \rm{km\ 
s}^{-1}$, which is well above the escape velocity. Note, that this maximal 
value
is attained close to the disk and that most of the gas in the outflow is a few 
$\rm{km\ s}^{-1}$ slower. The six protostars cover a range of luminosities 
between $0.22\ L_{\rm{\odot}}$ and $81.4\ L_{\rm{\odot}}$. We model the 
luminosities taking also the accretion onto the surface of the protostars into 
account which provides a significant part to the total luminosity in this stage 
of  star formation \citep[see][for details]{Offner2009}.\\
 The mid-plane gas number density $n_{\rm{g}}$, dust temperature 
$T_{\rm{d}}$, as well as the 3D magnetic field morphology of the considered 
snapshot are shown in Fig.  
\ref{fig:3DMag}. Investigating the magnetic field structure within the outflow 
lobes, we find that the toroidal magnetic field component clearly dominates over 
the poloidal component while the field outside the outflow lobe is hourglass 
shaped. Furthermore, we observed the occurrence of several shocks within the 
outflow locally leading to rather unordered gas motions and magnetic field 
orientations.

\section{Constraints to dust modeling}
\label{sect:DustModel}

The interpretation of observational polarization data strongly depends on the 
parameter of the considered dust grain model. While we know for sure that 
grains in the interstellar medium (ISM) are not spherically symmetric, their 
actual shape, 
composition and size distribution is uncertain. \\
A satisfactory model, the so called MRN model \citep[][]{1977ApJ...217..425M}, 
reproducing the galactic extinction curve, is a three parameter model with a 
power-law size distribution $n(a) \propto a^{-q}$ and a dust grain size range 
of 
$a \in [a_{\rm{min}}:a_{\rm{max}}]$ with values of $q=-3.5$, 
$a_{\rm{min}}=5\ \rm{nm}$, and $a_{\rm{max}}=250\ \rm{nm}$. In subsequent 
studies, the upper 
limit was extended to be of $\rm{\mu m}$ - size 
\citep[e.g.][]{2003ApJ...588..871C,2007ApJ...657..810D}. To account for 
characteristic extinction and absorption features, an ensemble of dust 
materials is used consisting of a mixture of carbonaceous (graphite) and  
silicate (olivene) materials \citep[e.g.][]{1995IAUS..163..355Z,Zhukovska2016}. 
Additionally,  in the case of IDG alignment
we considered ferromagnetic particles encapsulated in the dust grains enhancing 
paramagnetic alignment by a factor of $10^2$ 
\citep[][]{1967ApJ...147..943J,2007A&A...468L...9D,2009E&PSL.284..516B}. An 
oblate dust grain with an fixed aspect ratio represents a particularly 
promising 
approach for simulating the interstellar polarization and extinction data 
\citep[][]{1985ApJ...290..211L,1995ApJ...444..293K,1995ApJ...450..663H} so, 
here we use an average value of $0.5$.\\
Since, the different grain materials have unique dielectric and paramagnetic 
properties 
this also results in a unique alignment behavior. Analyzing the linear 
polarization and circular polarization shows that a higher alignment efficiency 
is to be expected for silicate grain while carbonaceous grains seem to 
remain unaffected by the presence of a magnetic field 
\citep[e.g][]{1976ApJ...207..126M,1986ApJ...308..281M}. Hence, we consider 
carbonaceous grains to be randomized, while silicate grains are partially 
aligned in our model \citep[][]{2003ApJ...588..871C,2007ApJ...657..810D}.\\
The dust properties required for the self-consistent dust grain heating 
calculations and polarization simulations are the efficiencies $Q$ of light 
polarization parallel ($Q_{\rm{||}}$) and perpendicular ($Q_{\rm{\bot}}$) to 
the 
grain symmetry axis \cite[see e.g.][]{1971MNRAS.153..279M}. The efficiencies 
can 
be pre-calculated by a numerical method representing the dust grains shape as 
an 
array of material specific discrete dipoles \citep[][]{2000ascl.soft08001D}. In 
order to remain consistent with the constraints of interstellar dust parameter 
we used the DDSCAT 7.2 code \citep[see][]{2013arXiv1305.6497D} to calculate 
values of $Q$ for an aspect ratio of $0.5$ in a regime of wavelength $\lambda 
\in [0.9\ \rm{\mu m}: 3\ \rm{mm}]$ and grain sizes limits of $a_{\rm{min}}= 5\ 
\rm{nm}$ and $a_{\rm{max}}=2\ \rm{\mu m}$. Here, the number of dipoles in 
our calculations ranges from $N = 171\,500 - 296\,352$ to remain in the 
numerical limit ($a < 0.05 \lambda \sqrt[3]{N}/|m|$) of the DDSCAT code, where 
$|m|$ is the complex 
refractive index. For the input to DDSCAT we use the refractory indices of 
\citep{1984ApJ...285...89D,1993ApJ...402..441L,2000AAS...197.4207W}. Dust grain 
with sizes with radii larger than $a >2 \mu m$ are outside the reach of DDSCAT. 
In order to overcome the numerical limit we combined existing data from 
DDSCAT with data obtained by the MIEX code \citep[][]{2004CoPhC.162..113W} 
using mie-scattering to smoothly extrapolate our dust model up to an upper 
cut-off radius of $a_{\rm{max}} = 200 \rm{\mu m}$. We consider the upper radius 
in our outflow environment to be larger than in the ISM and make the 
model with $a_{\rm{max}}=2\ \rm{\mu m}$ as our default mode.\\
The POLARIS RT simulations are performed with the cross-sections for a dust 
grain of average size with
\begin{equation}
	\overline{C}_{\rm{X}}=  \sum_{i}{\kappa_i \cdot 
\int_{a_{\rm{min}}}^{a_{\rm{max}}} \pi a^2 (Q_{\rm{X,i,||}}+Q_{\rm{X,i,\bot}}) 
n(a)R(a) da},
	\label{eq:avgExt}
\end{equation}
where $\pi a^2$ is the geometrical cross-section, $\kappa_i$ is the fraction of 
distinct dust grain materials, and $\overline{C}_{\rm{X}}$ stands for the cross 
sections of extinction ($C_{\rm{e}}$), absorption ($C_{\rm{a}}$) and scattering 
($C_s$), respectively. The same averaging is applied for the cross sections 
$\Delta C_{\rm{e}}$ for dichroic extinction, for thermal re-emission $\Delta 
C_{\rm{a}}$, and circular polarization $\Delta C_{\rm{c}}$ with
\begin{equation}
	\Delta \overline{C}_{\rm{X}}=  \sum_{i}{\kappa_i \cdot 
\int_{a_{\rm{min}}}^{a_{\rm{max}}} \pi a^2 
(Q_{\rm{X,i,||}}-Q_{\rm{X,i,\bot}})\sin^2(\vartheta) n(a)R(a) da}.
	\label{eq:avgPol}
\end{equation}
Here, the cross sections are weighted by the Rayleigh reduction factor $R(a)$ 
\citep[see e.g.][for details]{1968nim..book..221G, 2007JQSRT.106..225L}, to 
account for imperfect grain alignment. $R(a)=1$ corresponds to perfect 
alignment along the direction of the magnetic field and $R(a)=0$ to randomly 
orientated dust grains, respectively. The angle $\vartheta$ is defined by the 
direction of the incident light and the magnetic field direction. Consequently, 
no linear polarization emerges along a LOS parallel to the magnetic field 
direction since the dust grain would appear to be spherical \citep[see][for 
details]{Reissl2014}.

\section{Dust grain alignment}
\label{sect:align}
The alignment of the rotation axis of a dust grain parallel to the direction of the 
magnetic field lines is due to paramagnetic effects within the grain material 
itself \citep[e.g.][]{1951ApJ...114..206D,B1915}. However, in the ISM  
perfect alignment is suppressed by gas-dust collisions and the 
interaction with the local radiation field.\\
Here, we go beyond previous approaches in this field 
\citep[e.g.][]{1995ApJ...444..293K,2009ApJ...696....1D,2012A&A...543A..16P} and 
include the the classical imperfect 
Davis-Greenstein (IDG) alignment due to paramagnetic relaxation 
\citep[][]{1951ApJ...114..206D,1967ApJ...147..943J,1979ApJ...231..404P} with 
as well as the 
radiative torque alignment (RAT) due to radiation-dust interaction 
\citep[][]{1976Ap&SS..43..291D,1996ApJ...470..551D,1997ApJ...480..633D, 
2007AAS...210.7904H,2009ApJ...697.1316H} in the MC RT simulations. Additionally, we consider the 
randomization of dust grains caused by thermal fluctuations in the dust grain 
material for the grain alignment efficiency
\citep[see][]{1997ApJ...484..230L}.\\
The IDG alignment is mainly determined by the parameter
\begin{equation}
	\delta_{\textit{0}} = 2.07 \cdot 
10^{20}\frac{\textbf{B}^2}{n_{\textit{g}} T_{\textit{d}} 
\sqrt{T_{\textit{g}}}},
	\label{eq:delta}
\end{equation}
that represents an upper threshold for the dust grain alignment. The IDG 
accounts for the alignment of small dust grains because grains with an 
effective 
radius above $\delta_{\textit{0}}$ do no longer significantly contribute to the 
net polarization \citep[see][for details]{1967ApJ...147..943J}. Ferromagnetic 
inclusions can enhance the grain alignment efficiency by several orders of 
magnitude \citep[see e.g.][for review]{Andersson2015}. \\
Irregular dust grains are expected to scatter left handed and right 
handed circular light differently 
\citep[][]{1976Ap&SS..43..291D,2003ApJ...589..289W}. This additional radiative 
torque (RAT) increases the aliment efficiency 
\citep[][]{2007AAS...210.7904H,2007AAS...210.7901A}. The RAT alignment assumes 
dust grains to align efficiently when the angular velocity 
$\omega_{\rm{rad}}$ resulting from RATs becomes dominant over the angular 
velocity $\omega_{\rm{gas}}$ caused by random gas bombardment so that 
$\omega_{\rm{rad}} \geq 3\times \omega_{\rm{gas}}$. Consequently, RAT alignment 
is determined by the ratio of angular velocities. The minimum grain radius 
$a_{\rm{alg}}$ at which dust grains start to align is determined by:
\begin{equation}
	\left(\frac{\omega_{\rm{rad}}}{\omega_{\rm{gas}}}\right)^2 = \frac{ 
a_{\rm{alg}} \rho_{\rm{d}}}{\delta 
m_{\rm{H}}}\left(\frac{t_{\rm{gas}}}{(t_{\rm{gas}}+t_{\rm{rad}})n_{\rm{g}}k_{B} 
T_{\rm{g}}} \int Q_{\Gamma}(\epsilon)\lambda\gamma_{\rm{\lambda}} 
\overline{u}_{\rm{\lambda}} d\lambda \right)^2.
	\label{eq:omega}
\end{equation}
Here, $\rho_{\rm{d}}$ is the density of the dust grain material and 
$\overline{u}_{\rm{\lambda}}$ is the local mean energy density of the radiation field.
 The wavelength 
specific anisotropy factor $\gamma_{\rm{\lambda}}$ varies between 
$\gamma_{\rm{\lambda}}=1$ for unidirectional radiation and 
$\gamma_{\rm{\lambda}}=0$ for an isotropic radiation field. For details about 
the constant $\delta$ and  the  characteristic gas drag time $t_{\rm{gas}}$ and 
thermal emission drag time $t_{\rm{rad}}$, respectively,  we refer to
\cite{1997ApJ...480..633D}. The radiative torque efficiency depends 
$Q_{\Gamma}(\epsilon)$ on the angle $\epsilon$ between the predominant 
direction 
of radiation and the magnetic field direction and allows to calculate the 
characteristic dust grain size $a_{\rm{alg}}$ at which dust grains start to 
align. For further details see \cite{Reissl2016}.

\section{Radiative transfer calculations}
\label{sect:RT}

\begin{figure*}[]
	\begin{minipage}[c]{1.0\linewidth}
			\begin{center}
			  \includegraphics[width=1.0\textwidth]{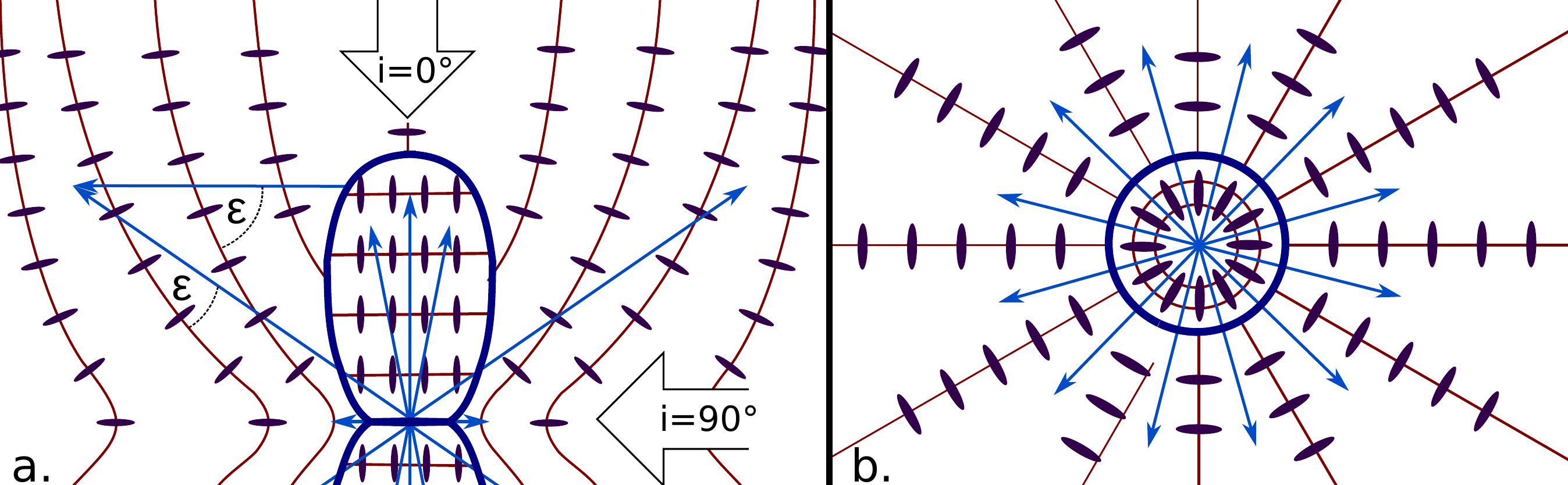}
			\end{center}
		\end{minipage}	
			
	\caption{Schematic illustration of the expected dust 
grain (dark blue ellipses) alignment behavior according to IDG and RAT 
alignment in the center plane perpendicular (panel a.) and parallel (panel b.) 
to the symmetry axis of the outflow lobes (blue). The angle $\epsilon$ is  
defined to be between the predominant direction of the radiation (light blue 
arrows) and the magnetic field lines (red). The white arrows indicate the 
definition of the inclination angle.}
\label{fig:SketchIDGRAT}
\end{figure*}
To create synthetic polarization maps we postprocess the data resulting from 
the MHD simulations discussed in Sect. \ref{sect:MHD} in a three step approach. 
At first, we update the initial MHD dust temperature in a Monte-Carlo (MC) 
simulation  
\citep[see][for 
details]{1999A&A...344..282L,2001ApJ...554..615B,Reissl2016} using the 
luminositis and position of the emerged cluster of protostars (see Sect. 
\ref{sect:MHD}).\\
Secondly, the anisotropy parameter $\gamma_{\rm{\lambda}}$ as well as the local 
energy density required by the RAT alignment is calculated in a second 
MC process. Here, we make again use of the method proposed in 
\cite{1999A&A...344..282L} using the path lengths and direction of all photons 
crossing a cell to obtain:
\begin{equation}
  \textbf{u}_{\rm{\lambda}} d \lambda =   \frac{\epsilon_{\rm{0}} }{c \Delta t 
V_{\rm{cell}}} \sum_{d \lambda} \frac{ l\cdot 
\textbf{k}}{\left|\textbf{k}\right|}.
\label{eq:LucyEnDens}
\end{equation}
With that, we can finally calculate the polarization maps. The wavelength regime considered in 
this paper covers the mid-IR to the far-IR, sub-mm and mm ($\approx 
20\ \rm{\mu m} - 3000\ \rm{\mu m}$). \\
A polarization pattern can also emerge because of scattering on dust grains. The quantity that quantifies the influence of scattering to the net polarization is the albedo $\alpha = C_{\rm{sca}}/C_{\rm{ext}} \in [0;1]$ where $\alpha = 1$ means that the polarization is completely dominated by scattering and $\alpha = 0$ stands for no scattering at all. Indeed, for the considered dust grain models with an upper cut-off radii of $a_{max} = 2\ \rm{mm}$ the albedo $\alpha \approx 0.25$ at a wavelength of $\lambda = 20\ \rm{\mu m}$. Hence, scattering can influence the polarization pattern especially near the disc region under such conditions. However, we start our investigation with an upper cut of radius of $a_{max} = 250\ \rm{nm}$ where $\alpha \approx 0.01$ at $\lambda = 20\ \rm{\mu m}$. Larger dust grains are just applied for synthetic polarization maps in the $mm$ regime of wavelength where $\alpha << 10^{-3}$ for all dust models. Hence, we neglect the influence of scattering within the scope of this paper.\\
The method of choice to represent the resulting polarization are the Stokes parameter $(I,Q,U,V)$. Here, $I$ stands for intensity, $Q$ and $U$ for linear polarization and $V$ for circular 
polarization. The radiative transfer in the Stokes formalism leads to a set of 
equations \citep[see][for details]{Martin1974} considering the dust 
grains as black body radiators that can be solved analytically. This allows
to calculate the contribution $(I',Q',U',V')$ of each cell
\citep[see][]{2002ApJ...574..205W} with the Planck function 
$B_{\rm{\lambda}}(T_{\rm{d}})$ along each path element $dl$ with

\begin{equation}
I'=\left(I+Q\right)e^{-n_{\rm{d}}l\left(C_{\rm{e}}+\Delta 
C_{\rm{e}}\right)}+n_{\rm{d}} 
lB_{\rm{\lambda}}(T_{\rm{d}})\left[C_{\rm{a}}+\Delta C_{\rm{a}} \cos(2\phi) 
\right],
\label{eq:trI}
\end{equation}

\begin{equation}
Q'=\left(I-Q\right)e^{n_{\rm{d}} l\left(\Delta 
C_{\rm{e}}-C_{\rm{e}}\right)}-n_{\rm{d}} 
lB_{\rm{\lambda}}(T_{\rm{d}})\left[C_{\rm{a}}+\Delta C_{\rm{a}} \cos(2\phi) 
\right],
\label{eq:trQ}
\end{equation}

\begin{equation}
U'=e^{-n_{\rm{d}} lC_{\rm{e}}}\left[U \cos(n_{\rm{d}}l\Delta C_{\rm{c}}) - V 
\sin(n_{\rm{d}}l\Delta C_{\rm{c}})\right] + n_{\rm{d}} l \Delta C_{\rm{a}} B_{\rm{\lambda}}(T_{\rm{d}})
\label{eq:trU}
\end{equation}
 and
\begin{equation}
V'=e^{-n_{\rm{d}} lC_{\rm{e}}}\left[U \sin(n_{\rm{d}} l\Delta C_{\rm{c}}) - V 
\cos(n_{\rm{d}} l\Delta C_{\rm{c}})\right].
\label{eq:trV}
\end{equation}
Here, $n_{\rm{d}}$ is the number density of the dust. The angle $\phi$ is 
between the direction of light polarization and the magnetic field lines. 
Note that because of the $\phi$ dependency circular polarization 
can only emerge in regions where each following magnetic 
field line and subsequently the preferential axis of grain alignment is 
non-parallel to the previous one along the LOS \citep[see e.g.][]{Martin1974}. 
The hourglass component in the outside region is such a magnetic field with rather 
parallel field lines along the LOS. As a result of this we get most of 
the circular polarization from within the outflow lobes in our RT 
calculations. Here, the maximum of the degree of circular polarization can 
amount up to $\pm 0.4\ \%$. As shown in \cite{Reissl2014} the pattern of 
circular polarization provides additional information to distinguish between 
different well ordered magnetic field morphologies. However, due to the chaotic 
nature of the helical field the results are 
inconclusive and do not allow to identify the helical component by any 
characteristic circular polarization pattern. Hence, we focus only on linear 
polarization in this paper. However, circular polarization still needs to be 
considered in the RT calculation because a permanent transfer from circular 
polarization ($V$ - parameter) to linear polarization ($U$ - parameter) and vise 
versa occurs. \\
The degree and 
orientation of linear polarization $P_{\rm{l}}$ is determined by 
\begin{equation}
P_{\rm{l}} = \frac{I_{\rm{p}}}{I} = \frac{\sqrt{Q^2+U^2}}{I},
\label{eq:Pl}
\end{equation}
where $I_{\rm{p}}$ is the polarized intensity. Its position $\chi_{\rm{Pl}}$ 
angle on the plane of the sky is defined by
\begin{equation}
\chi_{\rm{Pl}} = \frac{1}{2}\arctan \left( \frac{U}{Q} \right).
\label{eq:PolAng}
\end{equation}
For a more detailed description we refer to 
\cite{Reissl2014,Reissl2016}.

\section{Synthetic polarization maps}
\label{sec:results}
\begin{figure*}
	\centering
	 \includegraphics[width=1.0\textwidth]{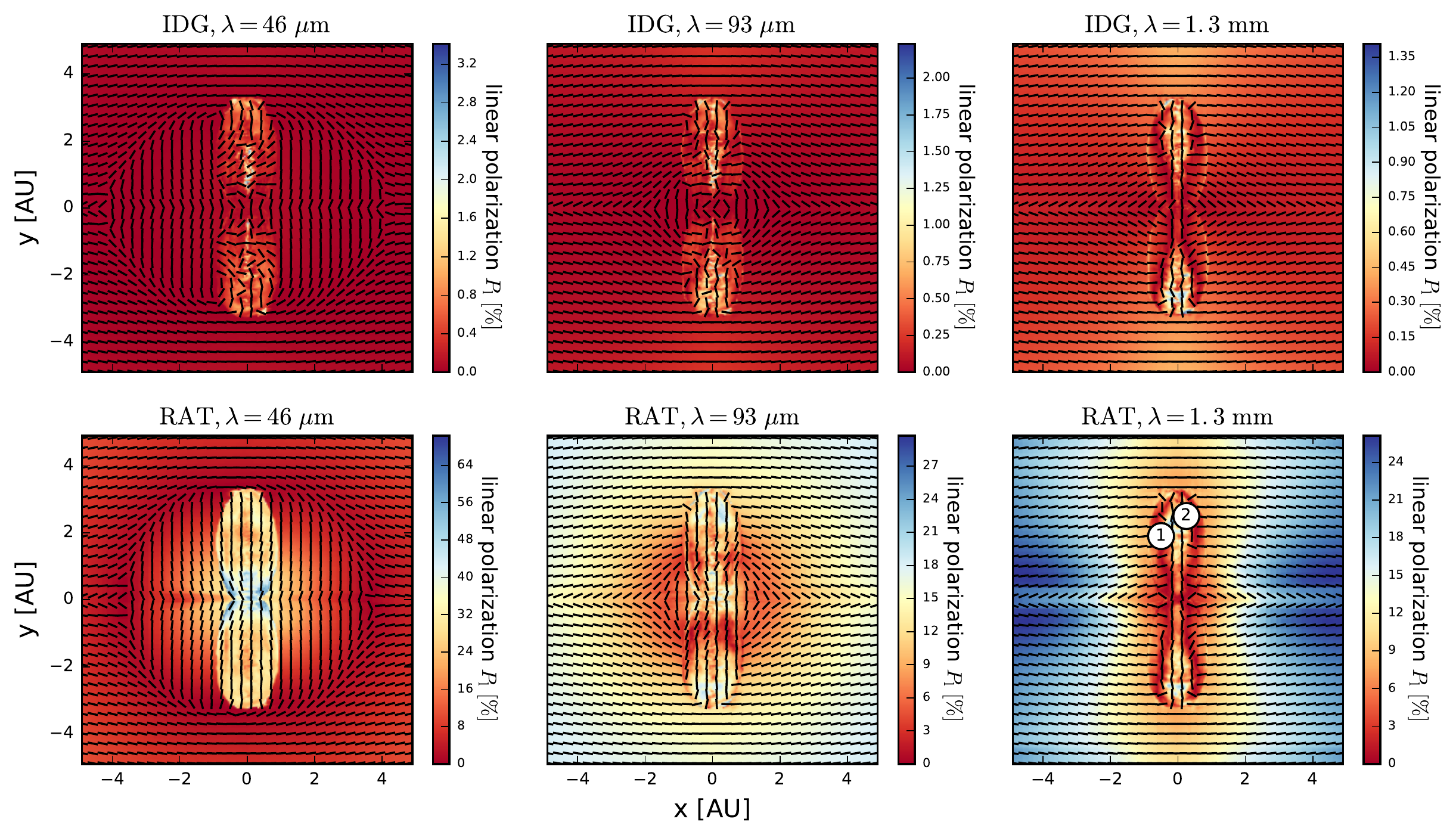}
	\caption{Maps showing the degree of linear polarization $P_{\rm{l}}$ 
(color coded) overlaid with normalized orientation vectors for the three 
distinct wavelengths of $\lambda = 46\ \rm{\mu m}$ (left column), $\lambda = 93\ 
\rm{\mu m}$ (middle column), and $\lambda = 1.3\ \rm{mm}$ (right column), 
respectively, at an inclination  angle of  $90^{\circ}$. The top row shows the 
maps considering IDG alignment while the bottom row shows the results 
considering RAT alignment. The numbers in the right bottom panel correspond to 
the LOS shown in Figs. \ref{fig:LOS01} and \ref{fig:LOS02}. Note, that only the
panels for $\lambda = 93\ \rm{\mu m}$ trace the helical magnetic field.} 
		\label{fig:ALLwave}

\end{figure*}

\begin{figure*}
\centering
	\includegraphics[width=1.0\textwidth]{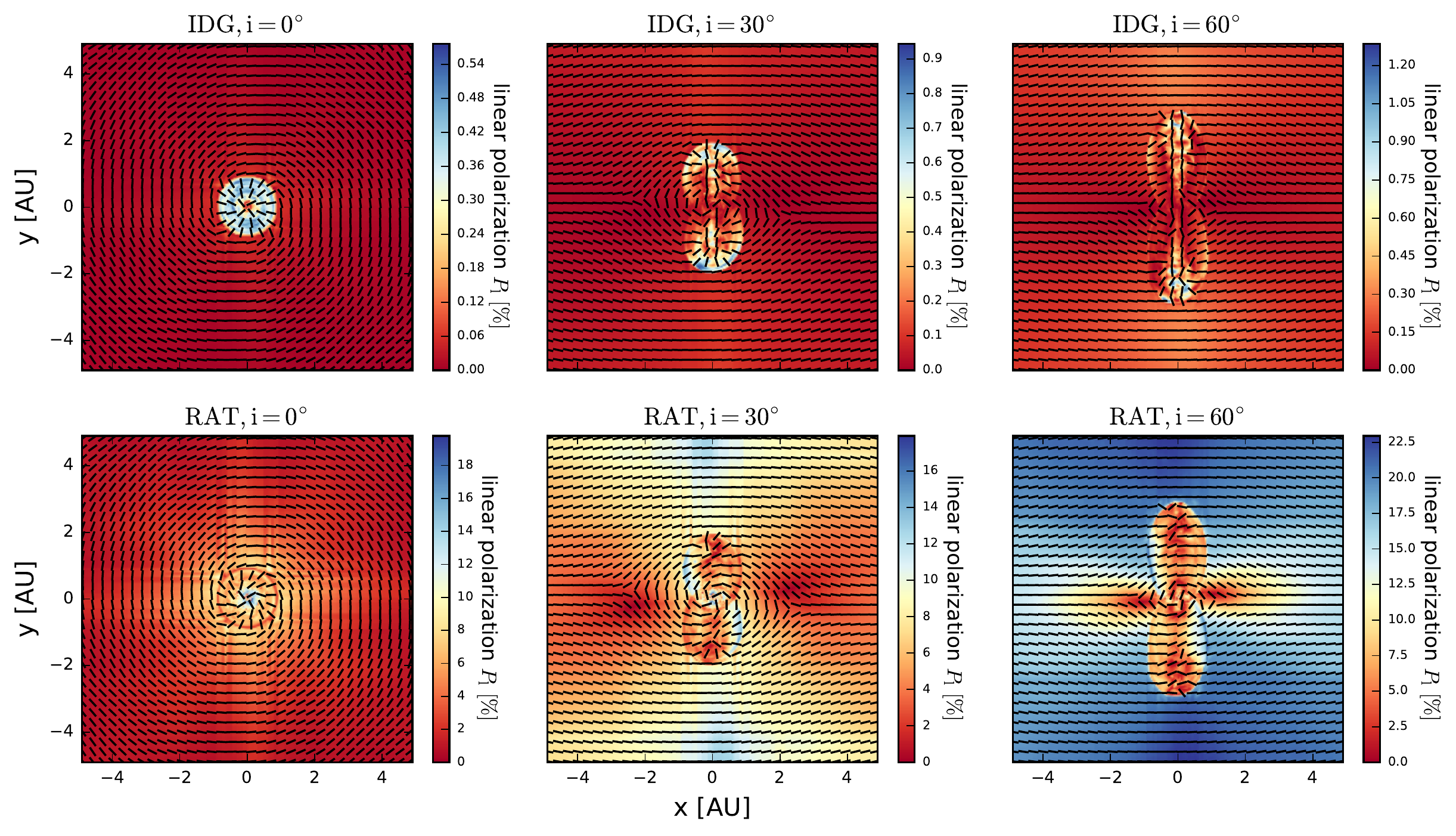}
	\caption{Maps showing the degree of linear polarization $P_{\rm{l}}$ 
(color coded) overlaid with normalized orientation vectors for a wavelength of 
$\lambda = 1.3\ \rm{mm}$  at the three distinct inclination angles $i = 
0^{\circ}$ (left column), $i = 30^{\circ}$, $i = 0^{\circ}$ (middle column), and 
$i = 60^{\circ}$ (right column), respectively. The top row shows the maps 
considering IDG alignment while the  bottom row shows the results considering 
RAT alignment. Note also that the polarization degree varies strongly strongly 
between the two alignment theories and for different wavelength.}
		\label{fig:ALLinc}
\end{figure*}

\subsection{Linear polarization by RAT and IDG alignment}
\label{sect:ALLwave}
In this section we investigate how the polarization pattern depends on grain 
alignment theory, inclination angle, and wavelength. Here, it is 
essential to estimate the expected contributions of different grain alignment 
theories to the net polarization separately. For the physical conditions 
present in the 
MHD simulation, both IDG and RAT alignment theory predict a similar behavior 
with regard to the preferential axis of grain alignment. However, the alignment 
efficiency and consequently the detectability of a polarization signal may 
differ significantly for different alignment theories. Furthermore, the two 
competing polarization mechanisms of dichroic extinction and thermal 
re-emission 
contribute both to linear polarization. Dichroic extinction dominates 
polarization in the UV, optical, and near-IR regime and leads 
to polarization parallel to the magnetic field direction. In contrast to 
dichroic extinction, thermal re-emission results in light polarization 
perpendicular to the magnetic field for a wavelength from the mid-IR to the 
mm. It is expected that in an intermediate regime of wavelengths both effects 
cancel each other out and that therefore the polarization vectors flip their orientation 
by $90^{\circ}$ in the transition. Hence, in order to determine the intermediate 
regime of wavelengths where dichroic extinction transitions to thermal 
re-emission we simulated synthetic polarization maps. The maps are calculated 
with an upper cut-off radius of $a_{\rm{max}} = 2\ \rm{\mu m}$ and cover a 
wavelength regime of $20\ \rm{\mu m} - 3\ \rm{mm}$. 
The disk is seen edge-on.\\
In Fig. \ref{fig:SketchIDGRAT} an illustration of the expected grain 
alignment is shown for a better interpretation of the following polarization 
maps. For an inclination of $90^{\circ}$ the resulting maps of linear 
polarization calculated with IDG and RAT alignment theory are shown in Fig. 
\ref{fig:ALLwave}. Although the IDG and RAT alignment theories are based on 
different physical principles, the resulting overall orientation of polarization 
vectors over wavelength are comparable.However, we note that the degree of 
linear polarization varies significantly between the two alignment theories with 
RAT leading to mach larger polarization values.\\
With increasing wavelength, the contribution of thermal re-emission to 
polarization becomes increasingly dominant over dichroic extinction. Since both 
competing polarization effects contribute in directions perpendicular to each 
other, it is possible that the net 
polarization vector flips by $90^{\circ}$ or is even canceled out.\\
The wavelength at which thermal re-emission starts to appear in the 
polarization maps for both IDG and RAT alignment is at $\lambda 
\simeq 40\ \rm{\mu m}$. The outer most regions of the polarization maps are 
less dense than the center and hence less affected by dichroic extinction. Therefore, 
polarization by thermal re-emission appears first in the outer part of the 
polarization maps an moves towards the center for longer wavelength. For a 
wavelength of $\lambda \simeq 46\ \rm{\mu m}$ the outflow lobe is still 
unaffected by the effect of flipping polarization vectors (Figs. 
\ref{fig:ALLwave} left column). Here, the outer regions of the polarization maps 
are dominated by thermal re-emission while the center regions are still 
polarized due to dichroic extinction. In the case of RAT alignment (Figs. 
\ref{fig:ALLwave} left bottom panel) this leads to a characteristic ring-shaped 
gap with a minimum degree of linear polarization where the contributions of 
dichroic extinction and thermal re-emission cancel out each other. For the IDG 
alignment (Figs. \ref{fig:ALLwave} left top panel) this effect is less 
pronounced and hardly detectable because of the overall low degree of linear 
polarization 
in that area.The IDG alignment is highly suppressed in high density and 
temperature regions 
(see Eq. \ref{eq:delta}) and would thus allow no conclusion about the 
underlying magnetic field morphology near the disk region.\\
At a wavelength of about $\lambda \simeq 93\ \rm{\mu m}$ (see Fig. 
\ref{fig:ALLwave} middle column) the outflow lobes are 
completely enclosed by the region of thermal re-emission while the 
polarization in the lobes itself results from dichroic extinction. This is due 
to the dust temperature of $T_{\rm{d}} \approx 30\ \rm{K} - 40\ 
\rm{K}$ at the surface of the outflow lobes (see Fig. \ref{fig:3DMag} middle 
panel) corresponding to a pack emission at $\lambda \approx 70\ \rm{\mu m} - 
97\ \rm{\mu m}$. The outside regions with $T_{\rm{d}} \approx 10\ \rm{K}$ 
contributes a neglectable amount of radiation at that wavelength regime. 
Consequently, the dust grains in front of the outflow lobes are illuminated by 
a strong background radiation emitted by the surface of the outflows and 
dichroic extinction dominates the polarization.\\
 Going to even longer wavelength for both RAT and IDG, 
thermal dust re-emission dominates the entire map at $\lambda 
\simeq 600\ \rm{\mu m}$ resulting to a polarization pattern comparable to that 
shown in Fig. \ref{fig:ALLwave} in the right columns. For larger wavelength on 
the polarization vectors are 
perpendicular to the projected magnetic field and their orientation remains 
rather constant up to the $\rm{mm}$ regime of wavelength.\\
As shown by detailed analysis (see Sect. \ref{sect:LOS}) even in the inner disk 
region linear polarization is completely due to thermal remission. Hence, in 
the 
maps considering IDG and RAT alignment the orientation of the polarization 
vectors represent the projected magnetic field morphology. In contrast to 
RAT alignment the helical component remains slightly more apparent at the tops 
and near the symmetry axis of the outflow lobes for IDG.


\subsection{Impact of inclination angle}
\label{sect:inc}
So far, we examined the polarization orientation and degree dependence on grain 
alignment theory and wavelength for a fixed inclination angle of $90^{\circ}$ 
between the outflow axis and the LOS. However, interpretation of polarization 
measurements are also influenced by projections effects. In this section we 
to investigate how the linear polarization pattern changes as a function of 
the inclination towards the observer. Due to the ambiguities in the 
mid-IR and sub-mm discussed in Sect. \ref{sect:ALLwave} and  with regards to 
the simulation of synthetic observations with the $ALMA$ telescope array we 
focus here on a wavelength of $\lambda = 1.3\ \rm{mm}$, where the polarization 
is purely due to thermal re-emission, i.e. the polarization vector is 
perpendicular to the field.\\
Fig. \ref{fig:ALLinc} shows polarization maps considering 
IDG alignment in comparison to RAT alignment for the three different 
inclination angles $i$ of $0^{\circ}$ (disk is face on), $45^{\circ}$,  and 
$90^{\circ}$, respectively. Again, note that the polarization degree differs by 
more than one order of magnitude. For IDG alignment and an inclination angle of $i = 
0^{\circ}$ (Fig. 
\ref{fig:ALLinc} top left panel), the LOS towards the center of the maps is 
parallel to the hourglass magnetic field  (see Fig. \ref{fig:SketchIDGRAT}). 
Since no linear polarization can occur in the case of a LOS parallel to the 
magnetic field (see Sect. \ref{sect:DustModel}), the linear polarization emerges 
completely in the interior of the outflow lobes and the disk component. With 
increasing inclination angle the contributions from the hourglass field start to 
dominate the overall orientation of linear 
polarization. For an inclination of $i = 45^{\circ}$ (see middle top panel 
of Figs. \ref{fig:ALLinc}) the polarization pattern begins to resemble the 
projected hourglass morphology. The helical component of the  magnetic field, 
however, remains apparent close to the symmetry axis and at the tips of 
the outflows, e.g. the orientation of the polarization vectors is predominantly 
vertical.\\
In the bottom row of Fig. \ref{fig:ALLinc} we show the linear polarization maps 
considering RAT alignment. For an inclination angle of $i = 0^{\circ}$ the 
results are qualitatively the same as the ones for IDG alignment.  However, 
with 
increasing inclination angle the contributions of the surrounding hourglass 
shaped magnetic field becomes even earlier dominant than for IDG alignment. 
Here, 
just the regions close to tips of the outflow lobes match the helical magnetic 
field component, i.e. the polarization vectors are vertically orientated.\\ 
When the subsequent magnetic field lines cross each other along the LOS, the 
polarized emissions of a given dust grain is canceled out by the emission of 
another one leading to an area of reduced linear polarization. This projection 
effect of crossing field lines along the LOS becomes apparent in the maps with 
RAT alignment (Fig. \ref{fig:ALLinc} bottom row) where two polarization holes 
(bottom middle panel) and two extra lobes of minimum of linear polarization 
(bottom right panel), respectively, become visible perpendicular to the symmetry axis of 
the outflow lobes. Here, these extra lobes are not a result of reduced dust 
density or temperature fluctuations, but are simply a projection effect and an indicator of the 
underlying hourglass-shaped field morphology \citep[see also][]{Reissl2014}.\\
The same effect can be observed for the polarization maps with IDG alignment in
the top middle panel and top right panel of Figs. \ref{fig:ALLinc}. However, 
the 
already low degree of linear polarization due to inefficient grain alignment 
outside the
outflow lobes makes these projection effects less relevant.\\
 Despite the additional ferromagnetic inclusions considered for the IDG 
alignment calculations, the maximum degree of linear polarization is of the 
order of a few per cent. This makes RAT alignment the relevant alignment 
process for observations in the presented outflow environment. Hence, recent 
efforts of \cite{Hoang2016} showed that it is possible to combine IDG and RAT 
alignment. However, we conclude that IDG is neglectable in the considered 
outflow environment and focus on RAT alignment alone in the following sections.

\begin{figure*}[ht]
		\centering
		\begin{minipage}[c]{0.99\linewidth}
		\centering
			 
		  \includegraphics[width=.49\textwidth]{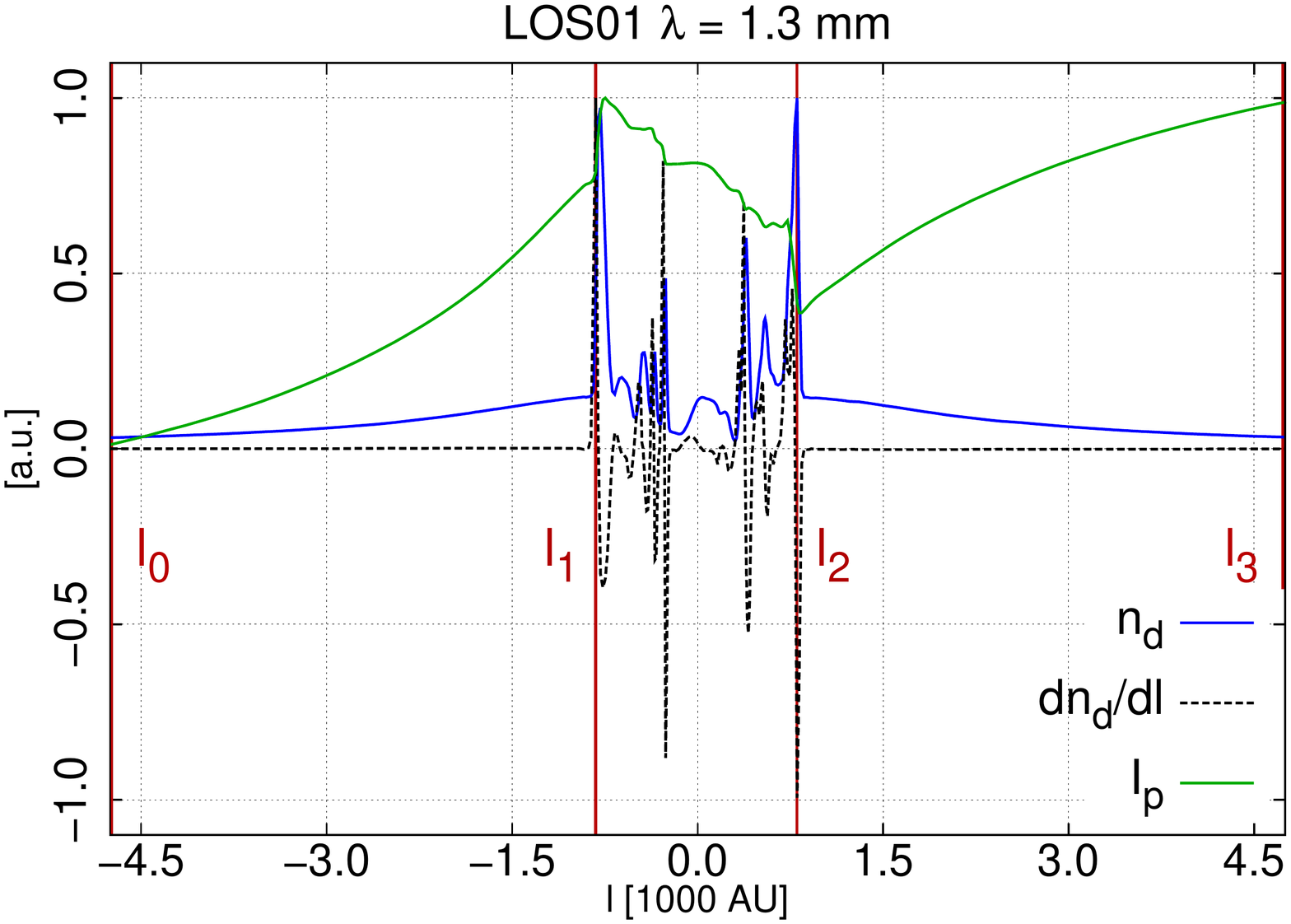}
		  \includegraphics[width=.49\textwidth]{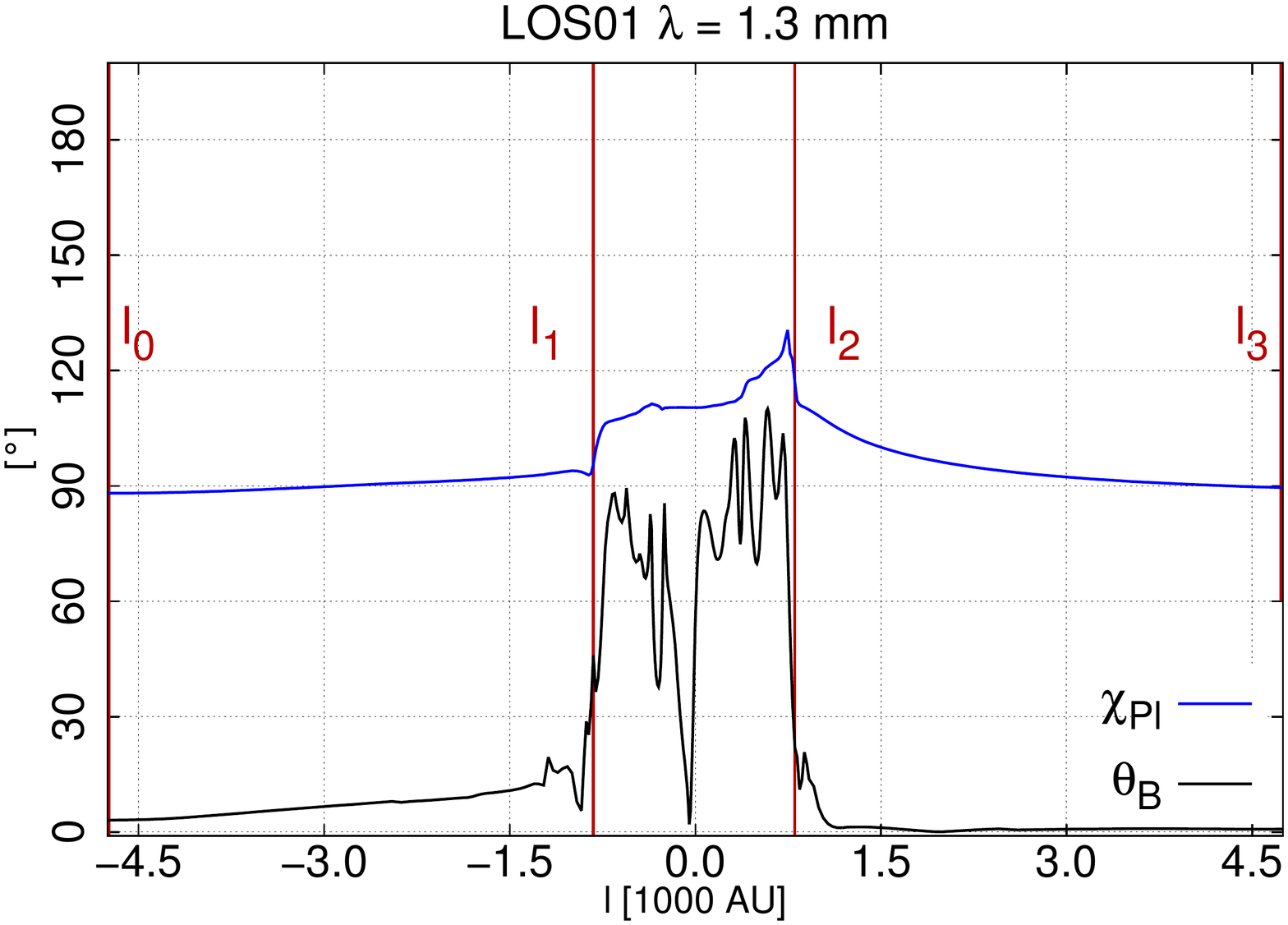}
			  
		  \vspace{-0.8cm}
		  \caption{Physical quantities along the a exemplary LOS 
  $ \# 1$ in the bottom right of Fig. \ref{fig:ALLwave}. The vertical red lines 
indicate the borders between the first outer region 
($l_{\rm{0}}-l_{\rm{1}}$), the interior of the outflow lobes 
($l_{\rm{1}}-l_{\rm{2}}$), and the second outer region 
($l_{\rm{2}}-l_{\rm{3}}$), 
respectively. Left panel: Normalized 
  values of polarized intensity $I_{\rm{p}}$ (green), dust number density 
  $n_{\rm{d}}$ (blue), and its first derivative $dn_{\rm{d}}(l)/dl$ (dotted 
  black). Note, that the polarized intensity $I_{\rm{p}}$ is plotted as a 
  cumulative value. Right panel: Orientation angles of linear polarization 
  $\chi_{\rm{P_l}}$ (blue) and magnetic field direction $\theta_{\rm{B}}$ 
(black). An orientation angle of $0^{\circ}$ corresponds to vertical 
polarization vectors in all figures shown in this paper.}
		  \label{fig:LOS01}
		\end{minipage}
		\begin{minipage}[c]{0.99\linewidth}
		\centering
		  \includegraphics[width=.49\textwidth]{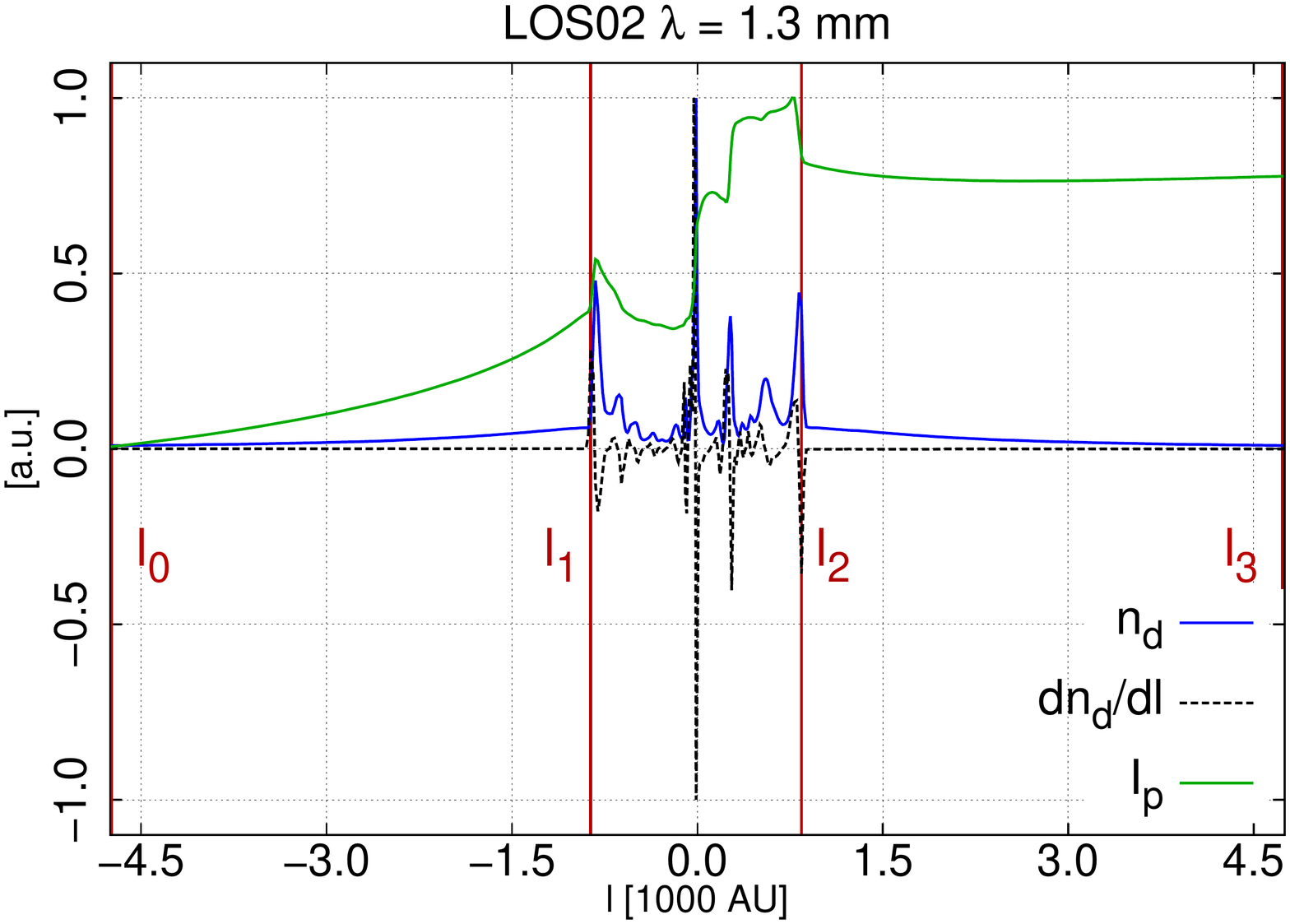}
		  \includegraphics[width=.49\textwidth]{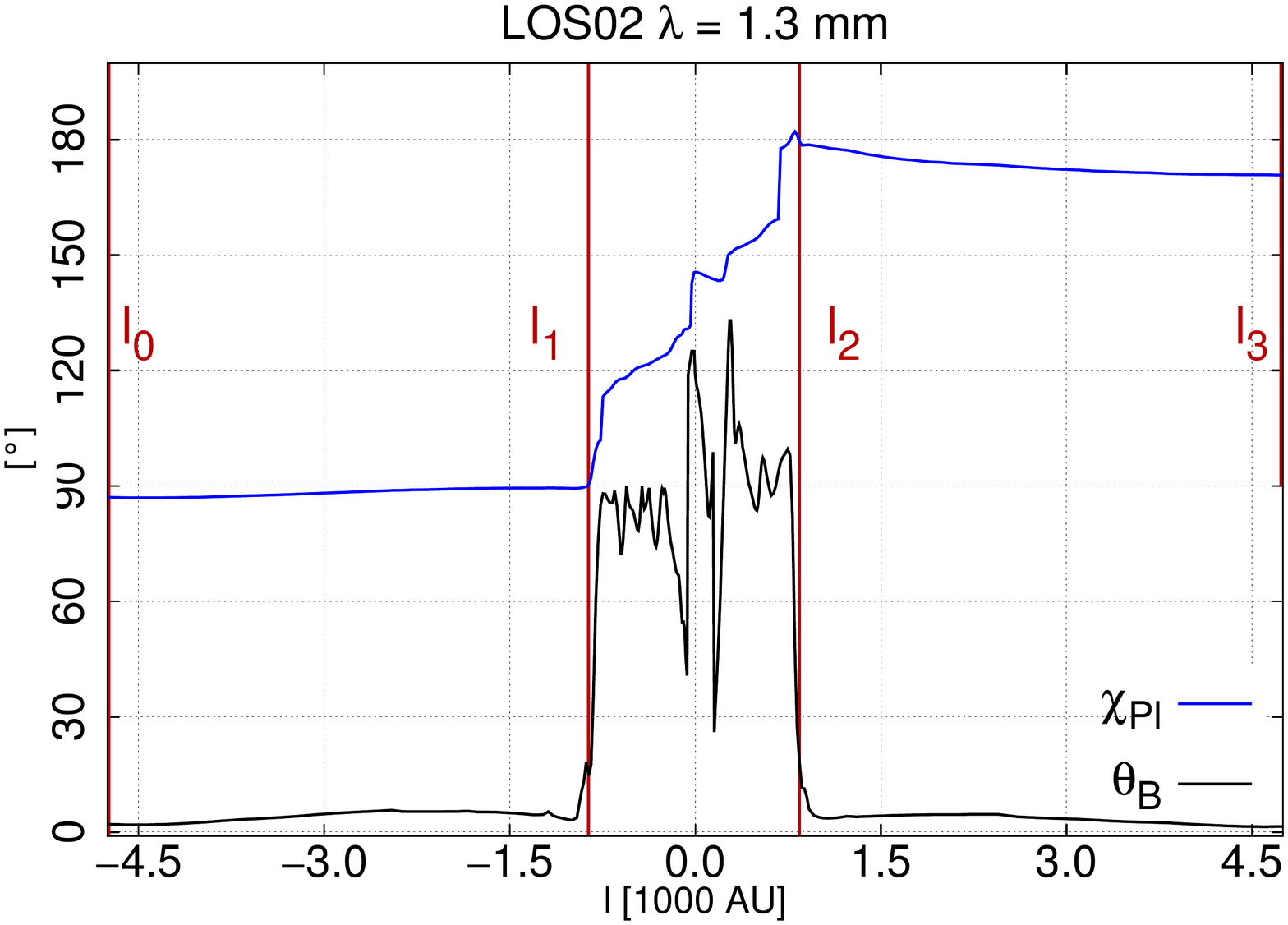}
		  \vspace{-0.8cm}
		  \caption{Same as Fig. \ref{fig:LOS01} along the LOS $ \# 2$ 
in the bottom right panel of Fig. \ref{fig:ALLwave}.}
		  \label{fig:LOS02}
		\end{minipage}

\end{figure*}
\begin{figure*}
	\centering
	\includegraphics[width=1.0\textwidth]{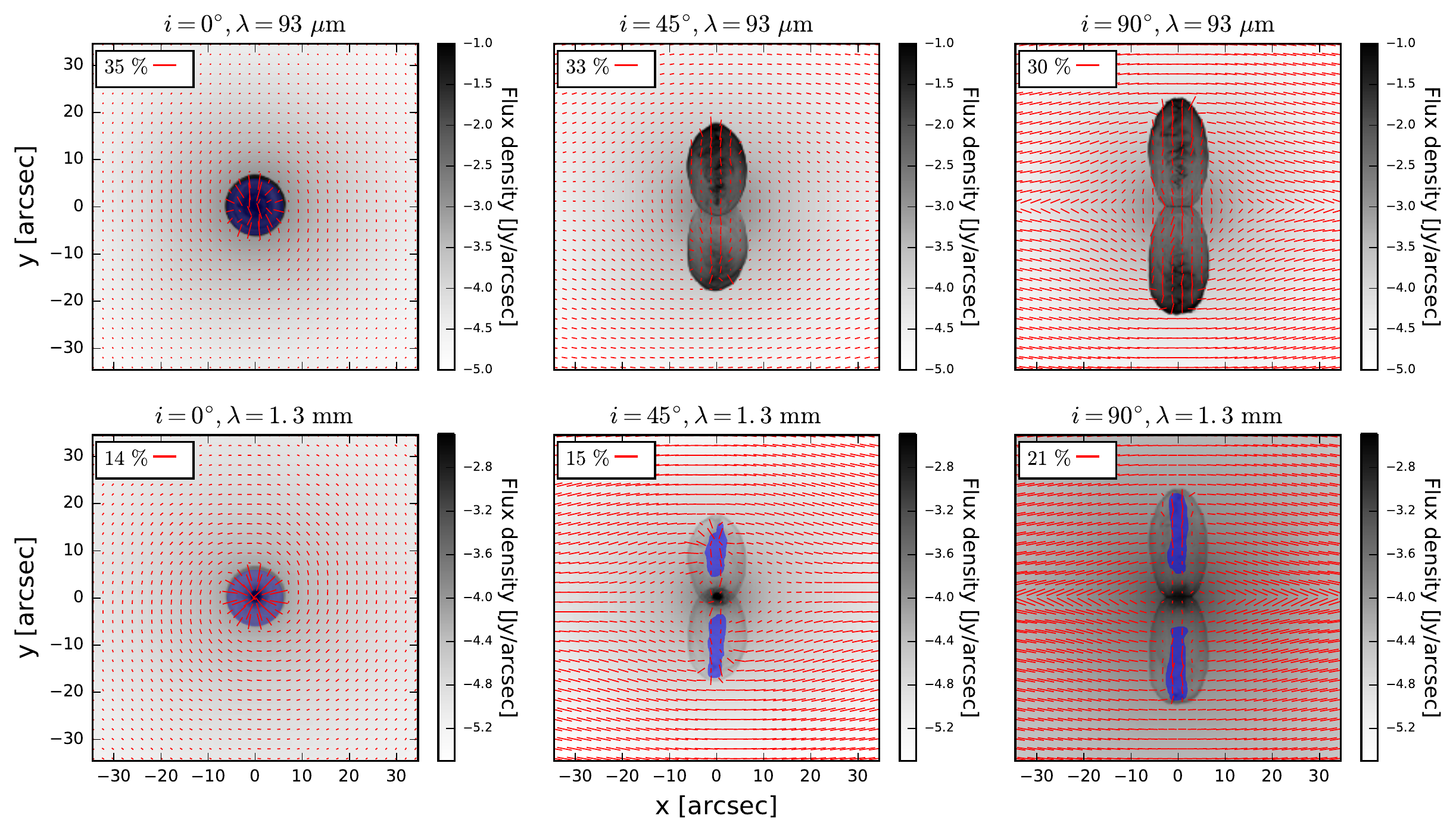}
\caption{Synthetic flux density maps (grey coded) with 
a wavelength of $\lambda = 93\ \rm{\mu m}$ (top row) and $\lambda = 1.3\ 
\rm{mm}$ (bottom row) and a distance of $140\ \rm{pc}$ overlaid with vectors of 
linear  polarization considering RAT alignment for inclination angles of 
$i=0^{\circ}$ (left column), $i=45^{\circ}$ (middle column), and $i=90^{\circ}$ 
(right column) and a maximal dust grain radius of $a_{\rm{max}}=2\ \rm{\mu m}$. 
The length of the vectors depends on the degree of linear polarization. For 
better comparison the colorbar is fixed in each row. The blue colored 
areas indicate where the polarization vectors represent the helical magnetic 
field structure of the outflow according to LOS analysis.}
   \label{fig:resLOS}
\end{figure*}

\section{Origin of linear polarization}
\label{sect:LOS}
In the previous sections we discussed maps of linear polarization on the basis 
of physically well motivated dust grain alignment theories and dust modeling. 
In these efforts it remained unclear to what extend the outflow lobes and the 
surrounding medium contribute to the synthetic maps of net polarization. 
Consequently, the synthetic polarization maps of Figs. \ref{fig:ALLwave} and 
\ref{fig:ALLinc} remain still ambiguous regarding the question of what component 
of the magnetic field is actually traced (hourglass \textit{in front} of the 
outflow lobes or \textit{in} the helical field).\\
This problem is even more severe for actual observations. Here, the spatial 
information of density and temperature as well as the magnetic field 
morphology in any observed astrophysical system gets lost due to the  
projection along a particular LOS. In contrast to observations, synthetic data 
by RT calculations allows to trace different LOS through the 3D MHD 
simulation and subsequently to determine the actual origin of polarization. In 
the following we focus on the origin and the actual 
detectability of polarization pattern characteristic for helical and hourglass 
magnetic field structures. Hence, we implemented an heuristic algorithm to 
analyze the polarization state of radiation along a distinct LOS $l$.\\
This automatized heuristic approach works in three steps:
\begin{enumerate}

\item Identification of the distinct regions inside and outside the outflow 
lobes. Here, we detect the bow shock of the outflow lobes by analyzing the first 
derivative of the gas number density $dn_{\rm{g}}/dl$ (Figs. \ref{fig:LOS01} 
and \ref{fig:LOS02} left panels in black dotted lines). This allows to 
distinguish between three regions along the LOS. The first outside region 
expands from $l_{\rm{0}}$ to $l_{\rm{1}}$, the outflows itself from $l_{\rm{1}}$ 
to $l_{\rm{2}}$, and the second outside region from $l_{\rm{2}}$ to 
$l_{\rm{3}}$ 
(indicated in Figs. \ref{fig:LOS01} 
and \ref{fig:LOS02} as red vertical lines).

\item Ray-tracing through the MHD simulation data in order to keep track of the 
accumulated polarized intensity $I_{\rm{p}}$ (Figs. \ref{fig:LOS01} 
and \ref{fig:LOS02} left panels in green lines), the orientation angle of 
linear 
polarization $\chi_{\rm{Pl}}$ (Figs. \ref{fig:LOS01} 
and \ref{fig:LOS02} right panels in blue lines) and the orientation 
angle $\theta_{\rm{B}}$ (Figs. \ref{fig:LOS01} 
and \ref{fig:LOS02} right panels in black lines) of the 
magnetic field with respect to the symmetry axis of the outflow lobes.

\item Determining origin of polarization by analyzing the largest increase in 
$I_{\rm{p}}$ as well as the relative orientation between magnetic field lines 
and linear polarization.

\end{enumerate}
With this simple but effective scheme, the actual origin of linear 
polarization can be determined. The first criterion is the increase or decrease, 
respectively, of polarized intensity $I_{\rm{p}}$ along each LOS. Here, it is 
sufficient to compare the accumulated polarized intensity at the points 
$I_p(l_{\rm{0}})$, 
$I_p(l_{\rm{1}})$, $I_p(l_{\rm{2}})$, and $I_p(l_3)$ to determine the area with 
the largest increase.\\
The second criterion is the resulting polarization angle with respect to the 
local magnetic field direction. Although the magnetic field direction in the 
outflow lobes is not well ordered it appears indeed rather regular in 
projection and can therefore be assumed to be perpendicular to the projected 
magnetic field direction in both outside regions.\\
We assume that the linear polarization originates from the interior of the 
outflow lobes when the largest increase in polarized intensity is inside the 
outflow lobe, and the orientation vector of linear polarization deviates from 
the projected helical field direction by less than $\pm 20^{\circ}$.\\
Here it needs to be emphasized, that this heuristic method is fine tuned to 
probe the 
outflow lobes in just this particular MHD simulation. The parameters of this 
heuristic approach are optimized by proper testing by minimizing the the false 
positive 
results. A manual evaluation of $75$ randomly chosen LOSs for each of the 
different inclination angles and wavelengths revealed that the accuracy of the 
correct detection of the origin of linear polarization is larger than $85\ 
\rm{\%}$.
Consequently, the areas in the polarization maps where linear polarization 
originates from the inside of the outflow lobes can be identified with high 
precision. However, the number of LOSs probing the helical field might actually 
be larger because of possible false negative detections.\\
Figs. \ref{fig:LOS01} and \ref{fig:LOS02} show the resulting plots of two 
exemplary LOSs corresponding to the positions in the right bottom panel of Fig. 
\ref{fig:ALLwave} at a wavelength of $\lambda = 1.3\ \rm{mm}$. In the left 
panel of Fig. \ref{fig:LOS01} the polarized intensity $I_{\rm{p}}$ emerges in 
the first outer region ($l_{\rm{0}}-l_{\rm{1}}$) and jumps to a maximum near 
the edge at $l_{\rm{1}}$ of the outflow lobe, decreases in the interior 
($l_{\rm{1}}-l_{\rm{2}}$) with a strong correlation to jumps in dust number 
density $n_{\rm{d}}$ and reaches its absolute maximum at the border of the grid 
at position $l_{\rm{3}}$.\\
In the first outside region ($l_{\rm{0}}-l_{\rm{1}}$) the orientation angle 
$\chi_{\rm{Pl}}$ (see \ref{eq:PolAng}) remains at an almost constant value of 
roughly $90^{\circ}$ with respect to the magnetic field orientation $\theta_{B}$ 
and increases 
slightly near the first edge ($l_{\rm{1}}$) of the outflow lobe as it is shown 
in the right panel of Fig. \ref{fig:LOS01}. In the interior 
($l_{\rm{1}}-l_{\rm{2}}$) the polarization angle remains again almost constant 
and trends back to $90^{\circ}$ as the radiation propagates towards the border 
($l_{\rm{3}}$) of the model space. In this case the interior of the outflow 
lobe is of minor influence to linear polarization with respect to degree and 
orientation. Clearly, the $LOS01$ does not probe the helical component of 
the magnetic field morphology but only the hourglass foreground. \\
Along the second LOS shown in Fig. \ref{fig:LOS02}, the polarized intensity 
$I_{\rm{p}}$ shows the same trend as in Fig. \ref{fig:LOS01} and the 
orientation angle $\chi_{\rm{Pl}}$ remains at around $90^{\circ}$ up to the 
point of $l_{\rm{1}}$ in the first outside region. However, in the center of 
the outflow lobe the accumulated amount of $I_{\rm{p}}$ rises to a global 
maximum an remains almost constant throughout the second outer region 
($l_{\rm{2}}-l_{\rm{3}}$). In contrast to $LOS01$, the orientation of linear 
polarization rotates by 
$90^{\circ}$ between $l_{\rm{1}}$ and $l_{\rm{2}}$. Here, the linear 
polarization matches the helical field component. In this case the resulting 
linear polarization probes the interior of the outflows and hence the helical 
magnetic field component.\\
The different polarization behavior between $LOS01$ and $LOS02$ may possibly 
attributed to a more unordered, and hence, less pronounced helical field 
component close to the borders of the outflow lobes. Furthermore, the gas 
temperature $T_{\rm{g}}$ is higher at the border. RAT alignment 
efficiency is inversely proportional $T_{\rm{g}}$ (see Eq. \ref{eq:omega}). 
This may also contribute the difference along $LOS01$ and $LOS02$.\\
Fig. \ref{fig:resLOS} shows maps of intensity  overlaid with vectors of 
linear polarization and different inclination angles for $\lambda=93\ \rm{\mu 
m}$ in comparison with maps for $\lambda=1.3\ \rm{mm}$. Here, we assumed only 
RAT alignment and a typical distance of a nearby 
star-forming region $140\ \rm{pc}$ 
\citep[e.g.][]{Preibisch1997,Torres2007,Mamajek2008}. Hence, the maps have 
a field of view of $69.1\ \rm{arcmin}$ with a 
resolution of $1024 \times 1024\ \rm{px}$ corresponding to $0.068 \times 0.068\ 
\rm{arcsec}$. Areas with a positive detection of the helical magnetic field 
component according to the LOS analysis presented in this section are marked in 
blue.\\
For the maps with $\lambda=93\ \rm{\mu m}$ (Fig. \ref{fig:resLOS} top row) 
the linear polarization traces the helical field component only for an 
inclination angle of $i = 0^{\circ}$. This is due the fact that linear 
polarization cannot emerge when the LOS is parallel to the magnetic field 
direction (see Sect. \ref{sect:DustModel}). Hence, the flip of the polarization 
vectors by $90^{\circ}$ for an inclination of $i = 45^{\circ}$ and $i = 
90^{\circ}$ is not due to the fact that we probe the inner helical field. We 
rather observe the outer hourglass field in extinction whereas in the 
surrounding it is probed in re-emission. Consequently, nor areas 
marked blue inn these panels.\\
In contrast to the maps with $\lambda=93\ \rm{\mu m}$, the synthetic 
observations at $\lambda=1.3\ \rm{mm}$  (Fig. \ref{fig:resLOS} bottom row) the  
polarization pattern here is due to thermal re-emission. Thus, any polarization 
vector flip by $90^{\circ}$ is an indicator only of a different magnetic field 
morphology. Indeed, the LOS analysis confirms, that the helical magnetic field 
morphology of the outflow can be well traced in the outflow center quite 
independent of inclination angle.\\
 The contribution of the disk region to the maps shown in Fig. 
\ref{fig:resLOS} is minuscule. For an inclination of $i=90^{\circ}$ the height 
of the disk is beneath the assumed resolution. With decreasing inclination the 
projected area of the disk increases. However, as the LOS analysis reveals, 
the polarization is completely dominated by the outflows or the outside 
regions but not for the disk. Hence, probing the 
magnetic field morphology in the disk itself seems to be impossible for the 
chosen MHD simulation and within the selected set of parameter presented in 
this paper.

\section{Grain size dependency}
\label{sect:GrainSize}
\begin{figure*}
	\centering
	\includegraphics[width=1.0\textwidth]{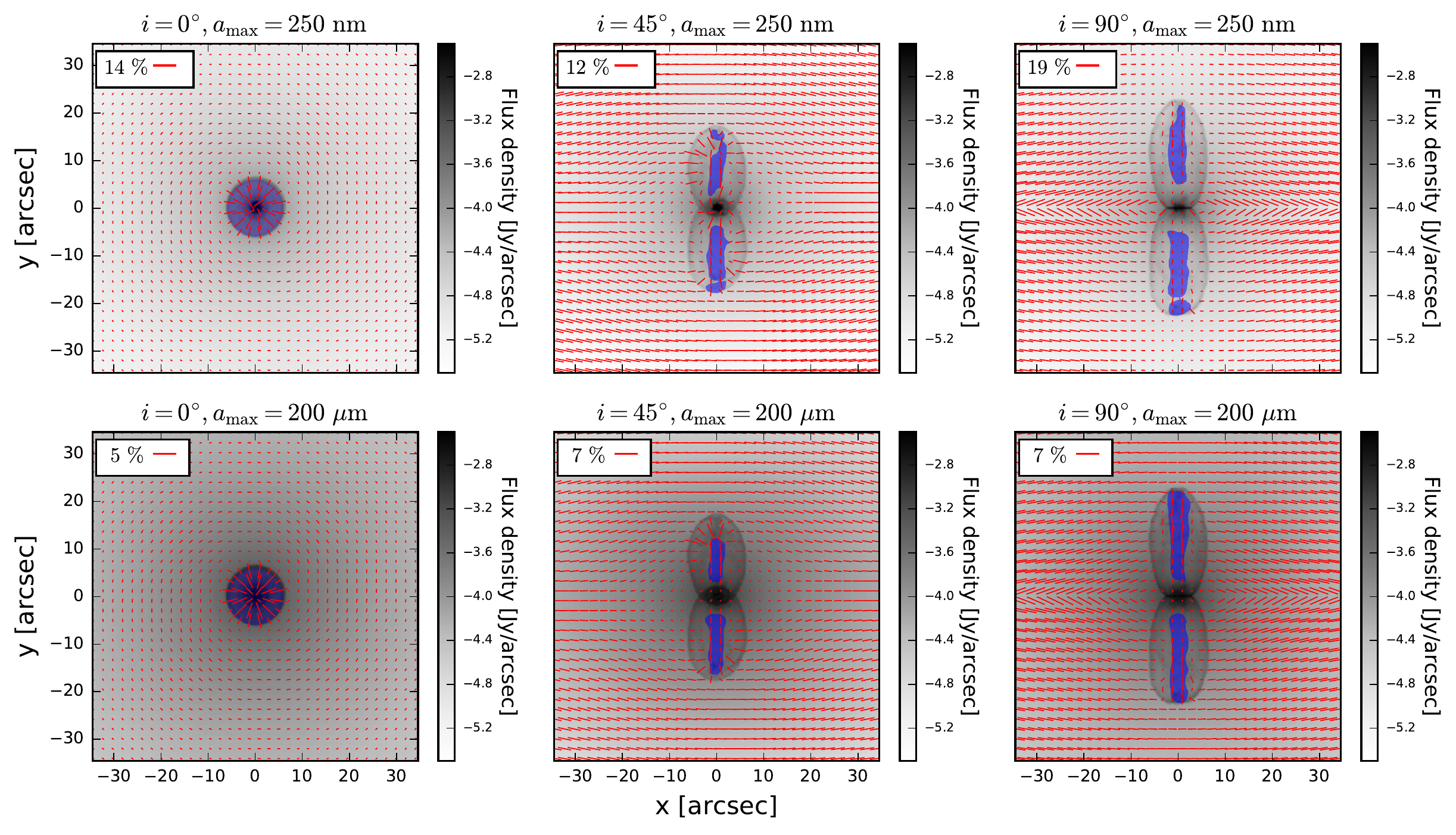}
\caption{Synthetic flux density maps (grey coded) 
with 
a wavelength of $\lambda = 1.3\ \rm{mm}$ and a distance of $140\ \rm{pc}$ 
overlaid with vectors of linear polarization 
considering RAT alignment for inclination angles of 
$i=0^{\circ}$ (left column), $i=45^{\circ}$ (middle column), and $i=90^{\circ}$ 
(right column) and maximal dust grain radii of $a_{\rm{max}}=250\ nm$ (top 
row) and $a_{\rm{max}}=200\ {\rm{\mu m}}$ (bottom row). The length of the 
vectors depends on the degree of linear polarization. For better 
comparison the colorbar is fixed between $F_{\rm{min}} = -4.0\ 
\log_{10}(\rm{Jy/arcsec})$ and 
$F_{\rm{max}} = -2.0\ \log_{10}(\rm{Jy/arcsec})$ in all plots. The blue colored 
areas indicate where the polarization vectors represent the helical magnetic 
field morphology of the outflow itself according to LOS analysis.}
   \label{fig:GrainDep}
\end{figure*}
Dust grains in the ISM are very likely to be of submicron size 
(see Sect. \ref{sect:DustModel}). In contrast to this, grains inside 
protoplanetary disks and outflows can 
differ significantly in size from grains in the diffuse ISM. Grain growth in 
the disk 
is expected to enrich the surrounding environment with dust grains of sizes up 
to $\lesssim 1\ \rm{mm}$ \citep[e.g.][]{Sahai2006,Hirashita2013}. The 
choice of upper dust grain size may significantly affect the 
resulting synthetic intensity and polarization maps and subsequently the 
predictions for future observational missions.
Hence, in this section we investigate the impact of the inclination angle 
together with how the upper dust grain size effects intensity and linear 
polarization.\\
The sizes of dust grains are most probably not evenly distributed in 
all regions. Larger dust grains coagulating in the disk are expected to get 
blown out along the outflow lobes. Near the tip of the outflow lobes, the dust 
grain sizes are re-processed towards smaller radii because of high temperature 
and pressure in this area. This complexity is beyond the scope of the current 
analysis and so we here follow a simpler approach with a constant upper cut-off 
radius of $a_{\rm{max}}$
throughout the model space. We 
pre-calculated dust cross sections for two additional dust models with distinct 
cut-off radii of $a_{\rm{max}} = 250\ \rm{nm}$ in accordance to the standard 
MRN model (see Sect. \ref{sect:DustModel}) and an extreme case with 
$a_{\rm{max}} = 200\ \rm{\mu m}$. We repeated the postprocessing of the MHD data 
for the RT  pipeline as described in Sect. \ref{sect:RT}.\\
Fig \ref{fig:GrainDep} shows the resulting map of intensity at a wavelength of 
$1.3\ \rm{mm}$ overlaid with the vectors of linear polarization dependent on 
inclination and the  different dust grain models. As for Fig. \ref{fig:resLOS}, 
we assume a distance of $140\ \rm{pc}$ to the outflow system. Areas with a 
positive detection of the helical magnetic field component according to the LOS 
analysis presented in the previous section are marked in blue. These findings 
agree independently of the applied dust grain model or inclination due to the 
fact that only the regions close to the symmetry axis of the outflow lobes are 
accessible by observations. The same applies for the orientation of linear 
polarization. The pattern of the polarization vectors shown in Fig. 
\ref{fig:GrainDep}  (see Fig. \ref{fig:resLOS} bottom row) is in good agreement 
with the results presented in Sect. \ref{sect:inc}. The overall polarization 
pattern calculated with an upper dust grain size of $a_{\rm{max}}=250\ \rm{nm}$ 
(Fig. \ref{fig:GrainDep} top row) and $a_{\rm{max}}=200\ \rm{\mu m}$ (Fig. 
\ref{fig:GrainDep} bottom row) are rather similar to each other. There 
are differences regarding intensity and degree of linear polarization. In 
contrast to the other maps with  $a_{\rm{max}}=250\ 
\rm{nm}$ synthetic images with an upper dust grain size of $a_{\rm{max}}=200\ 
\rm{\mu m}$ (Fig. \ref{fig:GrainDep} right column) show a higher intensity with 
an reduced degree of linear polarization. However, the finding that the helical 
magnetic field component within the outflow lobes should in principle be 
detectable for an wavelength of $\lambda = 1.3\ \rm{\mu m}$ holds independent 
of applied dust grain model (see also Fig. \ref{fig:resLOS}).


\begin{figure*}
		\centering
		\includegraphics[width=1.0\textwidth]{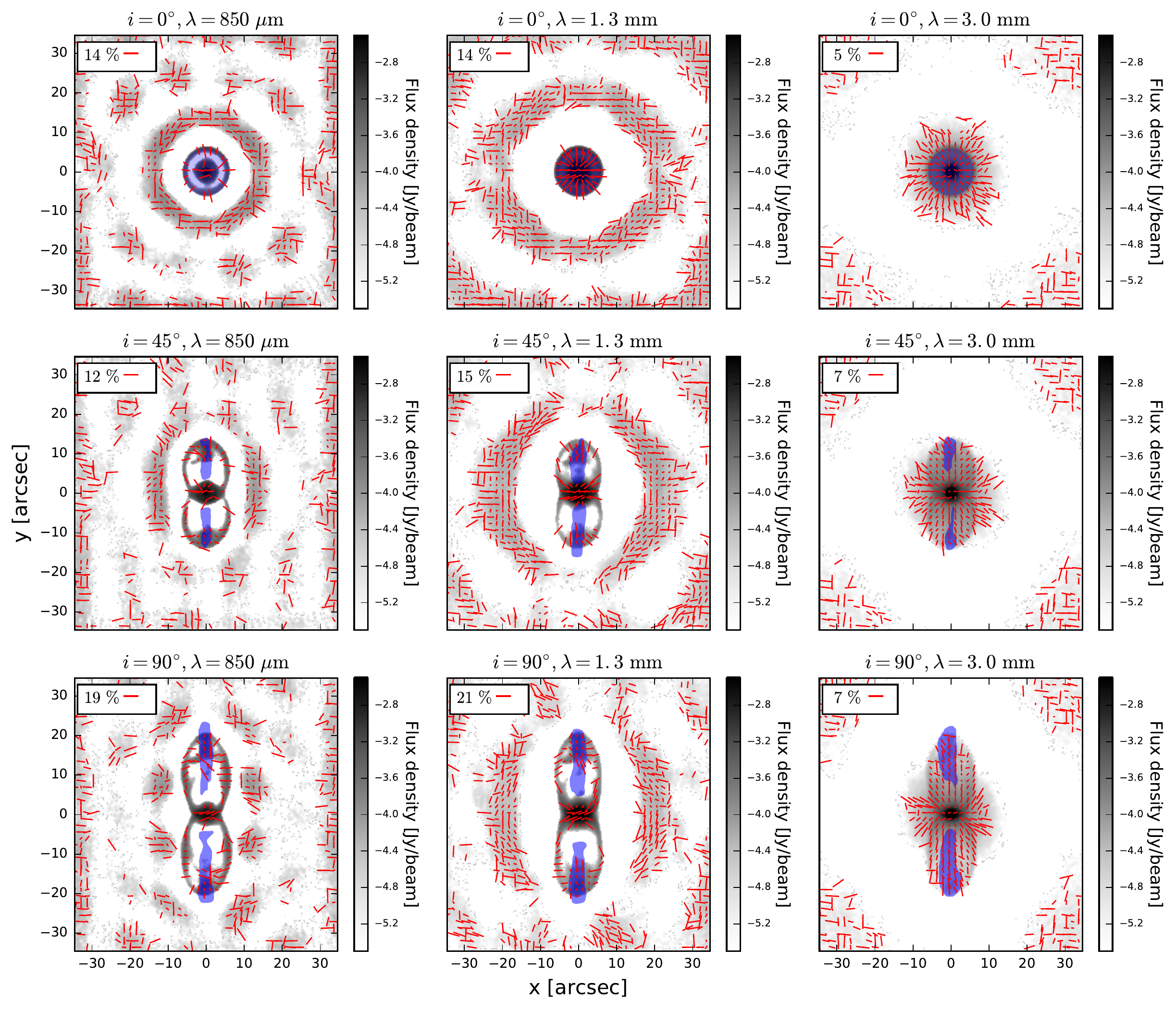}

\caption{Synthetic $ALMA$ flux density maps (grey coded) for  
a wavelength of $\lambda = 850\ \rm{\mu m}$ (left column), $\lambda = 1.3\ 
\rm{mm}$ (middle column), and $\lambda = 3.0\ \rm{mm}$ (right column) 
at a distance of $140\ \rm{pc}$ overlaid 
with vectors of linear polarization considering RAT alignment for inclination 
angles of $i=0^{\circ}$ (top row), $i=30^{\circ}$ (middle row), and 
$i=60^{\circ}$ (bottom row). The length of the vectors depends on the degree of 
linear polarization. For better comparison the colorbar is fixed between 
$F_{\rm{min}} = -4.0\ \log_{10}(\rm{Jy/beam})$ and $F_{\rm{max}} = -2.0\ 
\log_{10}(\rm{Jy/beam})$ in all plots. The blue colored areas indicate where the
helical magnetic field morphology is potentially be detectable accord LOS 
analysis.}
\label{fig:ALMA}

\end{figure*}

\section{Synthetic observations}
\label{sect:Observation}
In contrast to the ideal scenarios of the prevision sections, realistic 
observing conditions considering instrumental and atmospheric effects may 
further 
complicate 
the interpretation of polarization data. In this section we translate our 
synthetic 
intensity and polarization data into observational maps. Here 
the focus is on determining whether the areas where the helical field is 
dominating linear polarization presented in Sect's. \ref{sect:LOS} and 
\ref{sect:GrainSize}  would still be detectable by actual observations.\\
With regard to the limitations of wavelength without a flip of polarization 
vectors (see Sect. \ref{sect:ALLwave}) as well the expected field of view (see 
Fig. \ref{fig:resLOS} bottom row) of the post-processed MHD 
outflow simulation the $ALMA$ \citep[][]{2004AdSpR..34..555B} observatory 
provides the necessary equipment to constrain the observable parameter of 
linear polarization, and subsequently, that of the underlying magnetic field 
morphology. We calculate polarization and intensity data for three 
typical wavelength of $850\ \rm{\mu m}$, $1.3\ \rm{mm}$, and $3.0\ \rm{mm}$ 
for dust grains with an upper cut-off of radius of $a_{\rm{max}}=2\ \rm{\mu 
m}$ of the size distribution. We assume 
a mosaic observation with an object-observer distance of $140\ \rm{pc}$. The 
Stokes $I$, $Q$, and $U$ maps were separately processed, making use of the 
standard $ALMA$ reduction software $CASA$ \citep[][]{McMullin2007} with the 
$\texttt{simobserve}$ and $\texttt{clean}$ task. Here, we 
apply a resolution of $1\ \rm{arcsec}$, an observation time of five hours, and 
we included thermal noise as well as precipitable water vapor of $0.5\ \rm{mm}$ 
mimicking the atmosphere.\\
Initially, the observability of the outflow lobes in 
the resulting $I$, $Q$, and $U$ maps was  heavily disturbed by the very bright 
inner disk regions due to the dominating influence of the overall PSF. Since 
these regions are very compact, most of the emission is in the longest 
baseline. 
Hence, we limit the  
visibility (in the uv-plane) 
in order to take care of these bright disk regions. 
Finally, we combined the $I$, $Q$, and $U$ maps to create 
synthetic maps of intensity and linear polarization with Eq's. \ref{eq:Pl} and 
\ref{eq:PolAng}.\\
Fig. \ref{fig:ALMA} shows the resulting flux density maps overlaid with the 
scaled vectors of linear polarization for our simulated $ALMA$ data. 
In the synthetic observations with a wavelength of $\lambda = 850\ 
\rm{\mu m}$ (Fig. \ref{fig:ALMA} left column), the detectable intensity covers 
just the disk region and the very edges of the outflow lobes. This is 
related to the point spread function (PSF), dominated by the brightest 
contributions located in the inner disk and the followup cleaning process 
\citep[see][]{McMullin2007}. However, these detectable regions still coincide 
with the regions determined with the LOS analysis (see Sect. \ref{sect:LOS}). 
Although the $ALMA$ instrument seems to be barely suitable to detect a 
polarization signal emerging from our post-processed MHD simulation at $\lambda 
= 850\ \rm{\mu m}$,  it is in principle possible to detect a polarization 
signal from within the outflow lobe since the detectability increases even 
further for even longer wavelength. At a wavelength of $\lambda = 1.3\ \rm{mm}$ 
(Fig. \ref{fig:ALMA} middle column) the interior of the outflow lobe and 
subsequently the helical magnetic field morphology is partially observable for 
low inclination angles. Even with an increase in inclination angle up to 
$90^{\circ}$ the helical field would still be partly accessible by polarization 
measurements with $ALMA$. For $\lambda = 3.0\ \rm{mm}$ (see Fig. \ref{fig:ALMA} 
right column) the observable intensity completely covers the outflow lobes. 
The only exception is an inclination angle of $90^{\circ}$. Here, the blue area 
of the LOS analysis is is marginally larger then the intensity observable with 
$ALMA$. However, 
as Fig. \ref{fig:ALMA} shows in the right bottom panel a wavelength of $3.0\ 
\rm{mm}$ provides the 
optimal configuration to discriminate the helical magnetic field morphology 
embedded in the outflow lobes from the surrounding hourglass field. This is due 
to the largest area found with the LOS analysis coinciding with intensity and 
the polarization pattern minimally influenced by the PSF in the outflow lobes.

\section{Discussion}
\label{sect:disc}
\subsection{Dichroic extinction versus thermal re-emission}
The examination of multi-wavelength polarization measurements highlights one of 
the major obstacles for the interpretation from dust polarization measurements. 
The polarization effects of dichroic extinction and thermal re-emission 
contribute simultaneously to linear polarization with the preferential 
polarization axes perpendicular to each other. Both contributions to 
polarization can be calculated exactly (see Sect. \ref{sect:RT}). Here, the 
most relevant parameters are the cross section of dichroic extinction  
$\Delta C_{\rm{e}}$ and absorption $\Delta C_{\rm{a}}$. In the applied dust 
model of Sect. \ref{sect:DustModel} the cross $\Delta C_{\rm{e}}$ dominates 
toward shorter wavelength while for absorption $\Delta C_{\rm{a}}$ increase 
with wavelength. Hence, one can expect an flip of $90^{\circ}$ towards longer 
wavelength. \\
Dust temperature and number density also vary along the LOS and 
different regions of the system dominate the overall polarization calculation 
as a function of 
wavelength resulting in an additional offset of the projected magnetic field 
direction. As a result of this, a characteristic area where both effects cancel 
each other out manifests as a eminently obvious ring structure of minimal 
polarization shown in the left column of Fig. \ref{fig:ALLwave}. 
Polarization maps of that type are prone to misinterpretation.  
Several physical effects such as regions with a reduced dust density, a large 
amount of crossing magnetic field lines along the LOS, or a randomization of 
dust grain orientation by a high velocity stream 
\citep[e.g.][]{1998ApJ...502L..75R}, may also account for a reduction in the 
degree of linear polarization.\\
An even more ambiguous polarization pattern represents the result shown in the 
middle column of Figs. \ref{fig:ALLwave} and \ref{fig:resLOS}. A helical 
magnetic field morphology in the interior of the outflow lobe is expected from  
theoretical predictions. Hence, by comparing the alignment behavior shown in 
Fig. \ref{fig:SketchIDGRAT}  with the orientation vectors of linear polarization 
in the middle panels of Figs. \ref{fig:ALLwave} one might easily conclude to 
probe the distinct helical field of the overall magnetic field morphology. 
However, the LOS analysis introduced in Sect. \ref{sect:LOS} reveals a
different picture. The particular polarization map at a wavelength of $\lambda 
= 93\ \mu m$ matches just the hourglass component of the magnetic field 
morphology throughout the polarization map as a result of two different 
polarization effects. In the outer parts it is thermal reemission while close to 
the outflow lobes the polarization vectors switch by $90^{\circ}$ due to 
dichroic extinction. The helical magnetic field morphology can not be 
traced up to a wavelength of about $\lambda \gtrapprox 600\ \rm{\mu m}$.\\
These ambiguities become irrelevant with decreasing inclination angles. A 
polarization signal does not emerge from aligned dust grains when the LOS 
and the magnetic field direction are parallel (see Eq. \ref{eq:avgPol}). 
Consequently, the hourglass field cannot contribute to linear polarization with 
a LOS along the outflow direction (compare Fig. \ref{fig:SketchIDGRAT} a.). 
Hence, for low inclination angles the helical morphology of the magnetic field 
is not hidden by the hourglass component and can be identified unambiguously. \\
In \cite{Chapman13}, they investigate the correlation between the 
direction of the magnetic field lines in low-mass cores and the bipolar 
outflows and not detectable of differently shaped magnetic fields.
The direction and degree of linear polarization observed  is in good 
agreement with that in presented in Sect. \ref{sect:ALLwave}. Especially, the 
decreasing degree of linear polarization towards the outflow (see Figs. 
\ref{fig:resLOS} and \ref{fig:GrainDep}) is consistent with 
our result. This confirms the validity of the approach of creating synthetic 
polarization maps by combining dust combining dust continuum RT with dust grain 
alignment theory to create synthetic polarization maps.

\subsection{Alignment-specific polarization pattern}
The RT simulations reveal a marginal contribution of IDG alignment to the 
net polarization, making RAT clearly the dominant alignment mechanism in this 
outflow scenario. Since RAT alignment depends also on the alignment efficiency 
$Q_{\Gamma}(\epsilon)$ (see Eq. \ref{eq:omega}), where $\epsilon$ is the angle 
between radiation field and magnetic field, one could also expect to detect 
this 
characteristic in the resulting polarization pattern 
\citep[see][]{2011A&A...534A..19A,Reissl2016}. However, in the post-processing 
of the MHD simulation such an angle-dependent effect is not noticeable because 
of the local dust temperature distribution and its contribution to the overall 
radiation field. In contrast to a point-like source, the heated and dense 
surface 
of the outflow lobes acts like a light bulb that illuminates the model space 
from more than one direction (compare Fig. \ref{fig:SketchIDGRAT} left 
panel) 
and the radiation field is more diffuse as it would be for a
point-like source. Hence the angle-dependency of $Q_{\Gamma}(\epsilon)$ 
becomes less relevant.

\subsection{The role of dust grain size}
For the maps in Fig. \ref{fig:GrainDep} (upper row) we used the standard MRN 
model with an cut off of $a_{\rm{max}} = 250\ \rm{nm}$. In this model RT 
calculations showed that the maximum size of RAT alignment (see Sect. 
\ref{sect:RT}) exceeded the upper cut of radius in a considerable amount of 
regions in the outflow simulation ($a_{\rm{alg}} \geq a_{\rm{max}}$). One could 
thus expect a lower degree of linear polarization. However, due to the lack of 
larger dust grains the model with $a_{\rm{max}} = 250\ \rm{nm}$ 
(\ref{fig:GrainDep} top row) is optically thinner 
compared to other models with larger cut off radii. This results in less 
extinction and consequently flux 
and polarization remains similar to the model with $a_{\rm{max}} = 2\ \rm{\mu 
m}$ shown in Fig. \ref{fig:resLOS} (bottom row). The model with 
$a_{\rm{max}} = 200\ \rm{\mu m}$ is shown in Fig \ref{fig:GrainDep} (bottom 
row). Dust grains of maximal size are quite rare in the dust mixture because 
of the standard MRN power-law size distribution (see Sect. 
\ref{sect:DustModel}). However, the higher re-emission cross section 
$C_{\rm{a}}$ of larger dust grains at mm wavelengths compensates 
their lower abundance. This 
results in a higher flux for the model with $a_{\rm{max}} = 200\ \rm{\mu m}$ 
compared to the other models by a factor of $\approx 2$. Additionally, with 
increasing cut-off radii the polarization along the LOS becomes also rapidly 
dominated by larger dust grains. However, in the range of wavelength where 
$\lambda \approx a$ the cross section of re-emitted polarization has its 
minimum. Consequently, the peak value of linear polarization is reduced for a 
model with $a_{\rm{max}} = 200\ \rm{\mu m}$ by a factor between $2$ and $3$.

\subsection{Constraints to observational equipment}
The synthetic polarization maps show that the helical field morphology in the 
interior of the outflow lobes is not easily accessible by polarization 
measurements with aligned dust grains (see Figs. \ref{fig:resLOS} and 
\ref{fig:GrainDep}). This holds even more reconsidering the 
limitations of actual observational equipment. 
The instrument $HAWC+$ mounted on the airborne telescope 
$SOFIA$ \citep[][]{Dowell2013} is capable of linear polarization 
measurements. However, its field of view (between  $2.7 \times 1.7\ 
\rm{arcmin}$ and 
 $8.0 \times 6.1\ \arcmin{arcmin}$) and spectral coverage ($53\ \rm{\mu m} - 
214\ 
\rm{\mu m}$) makes it not suitable to observe the particular outflow scenario 
presented in this paper (see also Sect. \ref{sect:disc}). 
Additionally, the instruments limit of polarized intensity does not allow 
to probe the interior of the outflow. In order to utilize the full field of view 
in the available $HAWC+$ bands, our protostellar outflow object should also be 
in a distance between $10\ \rm{pc}$ and $40\ \rm{pc}$. However, there is no star 
forming region within such a distance \citep[][]{Preibisch1997}. Furthermore, as 
shown in Sect. \ref{sect:ALLwave}, the regime of wavelengths, where the 
transition of dominant polarization mechanism from dichroic extinction to 
thermalre-emission takes place, coincide with the $HAWC+$ bands. Consequently, 
the measurements with the $HAWC+$ instrument would be inconclusive with regard 
to the actually traced magnetic field direction. We performed an additional LOS 
analysis for all $HAWC+$ bands which reveal that the observed polarization 
pattern would completely represent the projected outside hourglass magnetic 
field morphology. The helical field inside the outflow lobe can not be detected 
within the $HAWC+$ bands. The detectability of the helical magnetic field 
morphology is just given in the far-IR,  sub-mm and mm regime of wavelength (see 
Sect.\ref{sect:LOS}).\\
In contrast to $SOPHIA/HAWC+$, the $ALMA$ telescope can 
actually probe the interior of the outflow lobes especially in the sub-mm 
and mm regime. Here, the limitations lie in the influence of 
the brightest regions in  the disk regions and their influence to the 
observability of the outflow lobes. While the outside regions can not completely 
be covered even at a wavelength of $\lambda = 3.0\ \rm{mm}$ the polarization of 
the outflow lobe, especially that emerging from the helical magnetic field 
component is accessible by $ALMA$ for that wavelength. However, we used an 
ideal non-turbulent MHD simulation for the RT calculations. Although ALMA 
observations can probe the interior of the outflow lobes this task may be 
challenged in more realistic environments with additional blending by 
turbulent motions.

\section{Summary and conclusions}
\label{sect:summ}
We presented synthetic polarization maps from the mid-IR to the mm 
wavelength regime of a post-processed MHD protostellar outflow simulation. The post-processing
was performed with the RT code POLARIS \citep[][]{Reissl2016}. Here, we considered different grain alignment theories, inclination angles, and dust models in order to constrain the parameters that 
allow to detect the helical magnetic field component in the outflow lobes 
embedded in a larger hourglass shaped field of the surrounding medium. \\
The conclusions of this study are as follows:
\begin{enumerate}
	\item With increasing wavelength the transition 
between dichroic extinction and thermal 
re-emission manifests itself in a flip of the orientation angles of 
$90^{\circ}$ in linear polarization. Additionally, areas where these transition 
takes place are depolarized and the magnetic field morphology is no longer 
accessible by observations. The polarization maps are completely dominated by thermal re-emission
at $\lambda \simeq 600\ \rm{\mu m}$ and the orientation in polarization pattern stays fixed. 
The low polarization degree in the outflows is in 
accordance with the findings of \cite{2011PASJ...63..147T} and \cite{Chapman13}.
	\item  We developed a heuristic method to identify the origin of the polarisation. We show that the helical magnetic field structure inside the outflow lobe is observable only close to the symmetry axis of the lobe and at the tip of the outflow. Outside these regions the polarisation emerges from the hourglass magnetic field structure in the foreground of the outflow.
	\item The alignment of dust grains does not result in polarization when 
the LOS is parallel to the magnetic field direction. This effect is independent 
of considered dust grain alignment theory and wavelength. However, this fact is 
of advantage in the particular case of the post-processed molecular outflows 
MHD simulation because it allows us to probe the interior of the outflow lobes 
for 
low inclination angles.
	\item Synthetic polarization maps have been calculated considering 
different grain alignment theories. The polarization is dominated by RAT alignment
that produces polarization degrees of a few $1\%$ to $\sim 10\%$ in agreement with 
observation. In contrast, the IDG alignment does not produce measurable polarization degrees. 
  \item 
	Probing the interior of the outflow lobes 
	depends on the maximum size of the dust grain distribution.
	We simulated polarization map with a power law size distribution considering 
	different upper  grain sizes.  A composition with 
larger grains leads to higher intensity but also a lower polarization and 
vise versa. However, the overall pattern of linear 
polarization seems to be independent of the cut-off radius. 
We expect the best observability for an upper cut-off in the order 
of $\approx 1\ \rm{\mu m}$.  
  \item From the observational point of view the best conditions to probe 
the interior of the outflow lobes is under inclination angles close to  
$0^{\circ}$ or $90^{\circ}$ for a wavelengths from the far-IR  
($\lambda \gtrapprox 600 \mu m$) to the mm regime. 	
  \item The interior of the outflow, i.e. the helical field structure, 
	cannot be probed with $SOPHIA/HAWC+$, since in the available bands the polarisation 
	is dominated by the hourglass field in the foreground of the outflows.
	
\item In contrast, ALMA observations should potentially allow to probe even the 
interior of 
the outflow lobes and subsequently to distinguish between the helical 
magnetic field in the outflow and the larger hourglass shaped field structure 
in the surrounding medium.
\end{enumerate}
As shown in this paper, the origin of the polarization 
remains ambiguous at best and cannot be easily inferred from observations alone.
However, progress is possible by creating physically well motivated synthetic 
polarization maps and designing methods of analysis allows to constrain the 
parameter of possible helical field detection for future observations. 

\begin{acknowledgements}
We wish to thank Gesa H. -M. Bertrang and Robert Brauer for useful discussions 
about RT and dust grain alignment. We also thank Eric Pellegrini and 
Thushara Pillai for their help with simulating synthetic $ALMA/CASA$ data.
For this project the authors S.R. and S.W. acknowledge the support of the DFG: 
WO 857/11-1. 
D.S. acknowledges funding by the DFG via the Sonderforschungsbereich SFB 956 
\textit{Conditions and Impact of Star Formation} as well as funding by 
the Bonn-Cologne Graduate School. The MHD simulations were performed at 
the supercomputer HLRB-II at the Leibniz Rechenzentrum in Garching.
We also acknowledge support from the Deutsche 
Forschungsgemeinschaft in the Collaborative Research Center (SFB 881) The Milky 
Way System (subprojects B1, B2, and B8) and in the Priority Program SPP 1573 
Physics of the Interstellar Medium (grant numbers KL 1358/18.1, KL 1358/19.2). 
RSK furthermore thanks the European Research Council for funding in the ERC 
Advanced Grant STARLIGHT (project number 339177).
\end{acknowledgements}

\bibliographystyle{aa}
\bibliography{./bibtex}
\end{document}